\def\kms{km~s$^{-1}$}

\def\ha{H$\alpha$}
\def\hb{H$\beta$}
\def\hg{H$\gamma$}
\def\hd{H$\delta$}
\def\lya{Ly-$\alpha$}

\def\oii{[O~{\sc ii}]}
\def\oiii{[O~{\sc iii}]}
\def\nii{[N~{\sc ii}]}

\def\neiii{[Ne~{\sc iii}]}

\def\hi{H~{\sc i}}
\def\hii{H~{\sc ii}}

\def\nhi{\mbox{$\sc N(\sc H~{\sc I})$}}

\def\caii{Ca~{\sc ii}}

\def\ci{C~{\sc i}}

\def\feii{Fe~{\sc ii}}

\def\mgi{Mg~{\sc i}}
\def\mgii{Mg~{\sc ii}}

\def\siii{Si~{\sc ii}}
\def\siiii{Si~{\sc iii}}

\def\oi{O~{\sc i}}

\def\ni{N~{\sc i}}

\documentclass[useAMS,usenatbib,epsfig]{mn2e}
\usepackage{rotating}
\usepackage{float}
\title[Absorbing Gas associated with a Galaxy Group]{Nature of the Absorbing Gas associated with a Galaxy Group at z$\sim$0.4\thanks{Based on observations collected at the European Organisation for Astronomical Research in the Southern Hemisphere under ESO programme(s) 096.A-0303.} }
\author[C\'eline P\'eroux et al.] {C\'eline P\'eroux$^1$\thanks{e-mail:celine.peroux@gmail.com}, Hadi Rahmani$^1$, Samuel Quiret$^1$, Max Pettini$^2$, Varsha Kulkarni$^3$, 
\newauthor
Donald G. York$^4$, Lorrie Straka$^5$, Bernd Husemann$^6$, Bruno Milliard$^1$\\
$^1$ Aix Marseille Universit\'e, CNRS, LAM (Laboratoire d'Astrophysique de Marseille) UMR 7326, 13388, Marseille, France.  \\
$^2$ Institute of Astronomy, University of Cambridge, Madingley Road, Cambridge CB3 0HA, UK.\\
$^3$ Dept. of Physics and Astronomy, Univ. of South Carolina, Columbia, SC 29208, USA.\\
$^4$ Dept. of Astronomy and Astrophysics and The Enrico Fermi Institute, University of Chicago, 5640 S. Ellis Ave, Chicago, IL 60637, USA.\\
$^5$ Sterrewacht Leiden, Leiden University, PO Box 9513, NL-2300 RA Leiden, the Netherlands.\\
$^6$ European Southern Observatory (ESO), Karl-Schwarzschild-Str.2, D-85748 Garching b. M\"unchen, Germany.\\
}

\begin{document}

\date{Accepted 2016 September 22. Received 2016 September 22; in original form 2016 July 29}

\pagerange{\pageref{firstpage}--\pageref{lastpage}} \pubyear{2016}

\maketitle

\label{firstpage}

\begin{abstract}
We present new MUSE observations of quasar field Q2131$-$1207 with a log \nhi=19.50$\pm$0.15 sub-DLA at z$_{\rm abs}$=0.42980. We detect four galaxies at a redshift consistent with that of the absorber where only one was known before this study. Two of these are star forming galaxies, while the ones further away from the quasar ($>$140 kpc) are passive galaxies. We report the metallicities of the HII regions of the closest objects (12+log(O/H)=8.98$\pm$0.02 and 8.32$\pm$0.16) to be higher or equivalent within the errors to the metallicity measured in absorption in the neutral phase of the gas (8.15$\pm$0.20). For the closest object, a detailed morpho-kinematic analysis indicates that it is an inclined large rotating disk with V$_{\rm max}$=200$\pm$3 km~s$^{-1}$. We measure the masses to be $M_{\rm dyn}$=7.4$\pm$0.4$\times$10$^{10}$ M$_{\odot}$ and M$_{\rm halo}$=2.9$\pm$0.2$\times$10$^{12}$ M$_{\odot}$. Some of the gas seen in absorption is likely to be co-rotating with the halo of that object, possibly due to a warped disk. The azimuthal angle between the quasar line of sight and the projected major axis of the galaxy on the sky is 12$\pm$1 degrees which indicates that some other fraction of the absorbing gas might be associated with accreting gas. This is further supported by the galaxy to gas metallicity difference. Based on the same arguments, we exclude outflows as a possibility to explain the gas in absorption. The four galaxies form a large structure (at least 200 kpc wide) consistent with a filament or a galaxy group so that a fraction of the absorption could be related to intra-group gas. 
\end{abstract}

\section{Introduction}
%%%%%%%%%%%%%%%%%%%%%%%%%%%%%%%%%%%%%%%%%%%%%
%%%%%%%%%%%%%%%%%%%%%%%%%%%%%%%%%%%%%%%%%%%%%

One of the key unknowns in the study of galaxy evolution is how galaxies acquire their gas and how they exchange this gas with their surroundings. Since gas, stars, and metals are intimately connected, gas flows affect the history of star formation and chemical enrichment in galaxies. Accretion is required to explain some of the basic observed properties of galaxies including the gas-phase metallicity \citep{erb06a}. Moreover, galaxies are believed to interact with the intergalactic medium (IGM) by pervading it with hydrogen ionising photons and by injecting heavy elements formed in stars and supernovae through supersonic galactic winds. Indeed, observations of the IGM indicate significant quantities of metals at all redshifts \citep{pettini03,ryanweber09,dodorico13,shull14a,becker15}. The presence of these metals is interpreted as a signature of strong galactic outflows in various models \citep{aguirre01,oppenheimer06}. Hydrodynamical simulations provide predictions of the physical properties of these gas flows \citep{keres05,brook11,keating15}. In recent years, much attention has been focused on the circumgalactic medium (CGM), a loosely defined term that describes the gas immediately surrounding galaxies over scales of $\sim 300$ kpc \citep{shull14b}. The CGM is at the heart of these physical processes. Therefore study of the CGM is crucial for understanding both the inflows of gas accreting into galaxies and the outflows carrying away the energy and metals generated inside galaxies. 

Outflows are commonly probed by the presence of interstellar absorption lines from cool gas blue-shifted by hundreds of km~s$^{-1}$ relative to the systemic velocities of the background galaxies \citep{shapley03,steidel10}. Strong \mgii\ absorbers in particular \citep{martin12,schroetter15} have been observed to extend out to 100 kpc along the galaxies' minor axes \citep{bordoloi11}, a fraction of which could be associated with galactic winds. Outflows are ubiquitous in galaxies at various redshifts \citep{pettini01,pettini02,cabanac08,heckman15}. Interestingly, the circumgalactic gas has also been probed in emission by \citet{steidel11} who stacked narrow-band images of z$\sim$2-3 galaxies and revealed diffuse Ly-$\alpha$ haloes extending to 80 kpc. More recently, the GTO (Guaranteed Time Observations) MUSE team used a 27-hr deep field to report Ly$\alpha$ haloes in individual  emitters down to a limiting surface brightness of $\sim$10$^{-19}$ erg/s/cm$^2$/arcsec$^2$ with a scale length of a few kpc \citep{wisotzki16}. 

While observational evidence for outflows is growing, direct probes of infall are notoriously more difficult to gather. So far, the accretion of cool gaseous material has been directly observed only in the Milky Way in high velocity clouds in 21 cm emission at distances of 5--20 kpc \citep{lehner11, richter14}. At larger distances, nearby spirals exhibit both extraplanar HI clouds and morphological disturbances, which may be attributed to gas infall \citep{sancisi08}. However, the emission from these diffuse structures is difficult to map at high redshift. Nevertheless, detections suggestive of cool gas inflows have recently been reported in a few objects \citep{rubin11,martin12,bouche13,diamond16}.

A powerful tool to study the CGM gas is offered by absorption lines in quasar spectra. We have initiated a novel technique to examine this gas in absorption against background sources whose lines of sight pass through the CGM of galaxies using 3D spectroscopy \citep{bouche07,peroux11a}. Over the past few years, we have demonstrated the power of this technique for studying the CGM using VLT/SINFONI by successfully detecting the galaxies responsible for strong \nhi\ absorbers at redshifts $z \sim 1$ and $z \sim 2$ \citep{peroux11b,peroux12,peroux13,peroux14,peroux16}. These detections have enabled us to map the kinematics, star formation rate, and metallicity of this emitting gas, and to estimate the dynamical masses of these galaxies. Out of our 6 detections for absorbers with known \nhi, we find evidence for the presence of outflows in two of them, while three are consistent with gas accretion. The remaining system at z$\sim$2 is poorly constrained  \citep{peroux16}.

Having demonstrated the power of 3D spectroscopy for study of the CGM at high-$z$ \citep{peroux16}, we now extend the technique at low-redshift with the MUSE optical spectrograph. Here, we present results from new observations of a sub-DLA at z$_{\rm abs}$=0.42980. The manuscript is organised as follows: Section 3 presents the ancillary and new observations of the absorber and the quasar field. Section 3 shows the analysis performed on the new MUSE and ancillary observations presented here. Finally, in section 4, we explore different scenarios to explain the gas seen in absorption in relation with the objects observed in the field. Throughout this paper we adopt an $H_{0}=70$~\kms~Mpc$^{-1}$, $\Omega_{\rm M}=0.3$, and $\Omega_{\rm \Lambda}=0.7$ cosmology. 
%--------------

\section{Observations of the Q2131$-$1207 Field}

\subsection{Quasar Spectroscopy: Absorption Properties}

\begin{figure}
\hspace{-0.8cm}
\includegraphics[angle=0,width=10.cm]{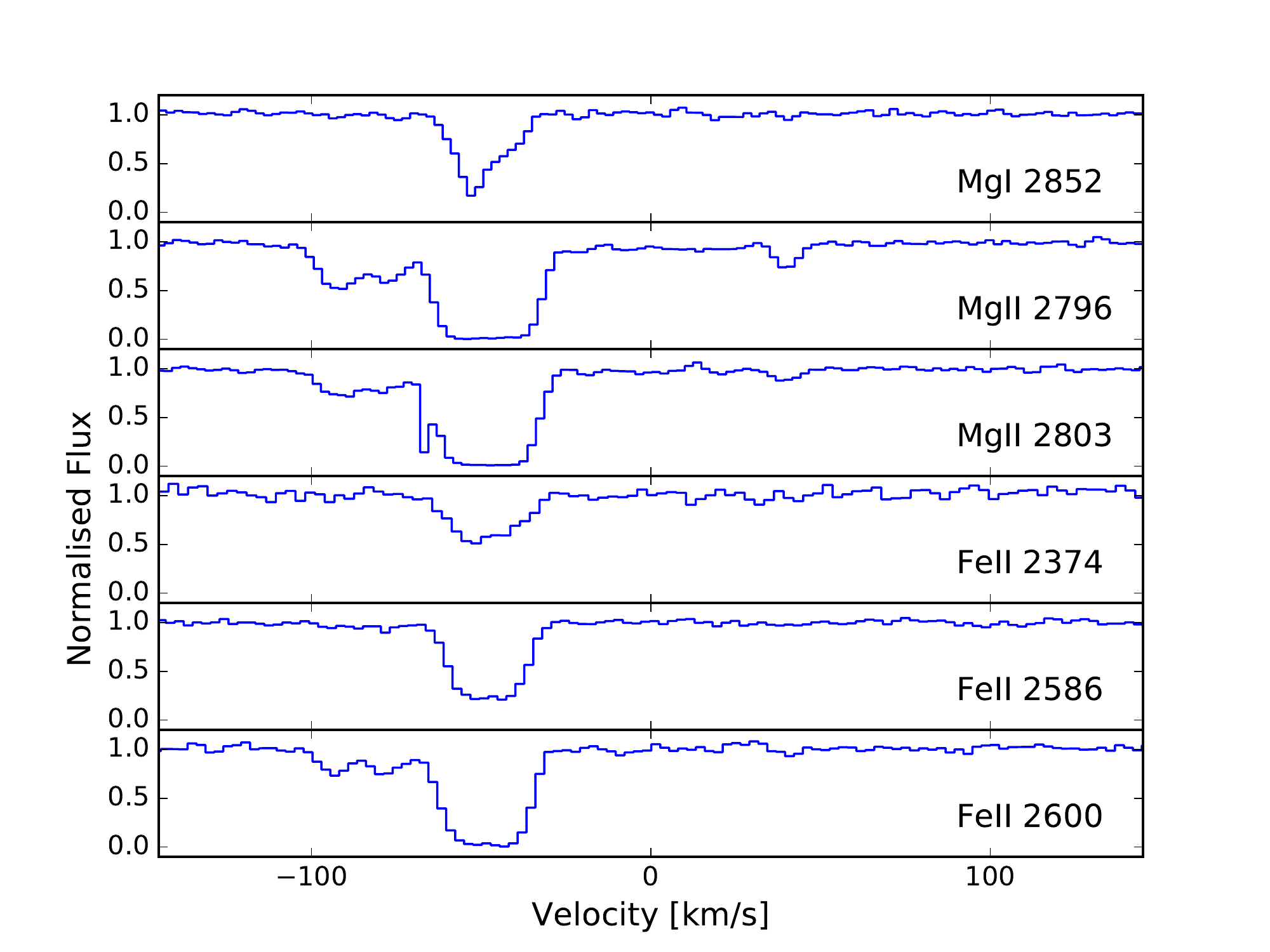}
\caption{{\bf Absorption profiles.} Excerpts from the high-resolution Keck/HIRES spectrum of the quasar Q2131$-$1207 showing absorption lines of \feii, \mgii\ and \mgi. The zero velocity component is set to the redshift of galaxy "a", z$_{\rm gal}$=0.43005. Most of the absorption appears to lie bluewards of this systemic redshift. We note however one weak component at $\sim +50$ km~s$^{-1}$ which is only seen in the strongest transitions (i.e. \mgii). }
\label{f:HIRES}
\end{figure}

\begin{figure}
\includegraphics[angle=0,scale=0.43]{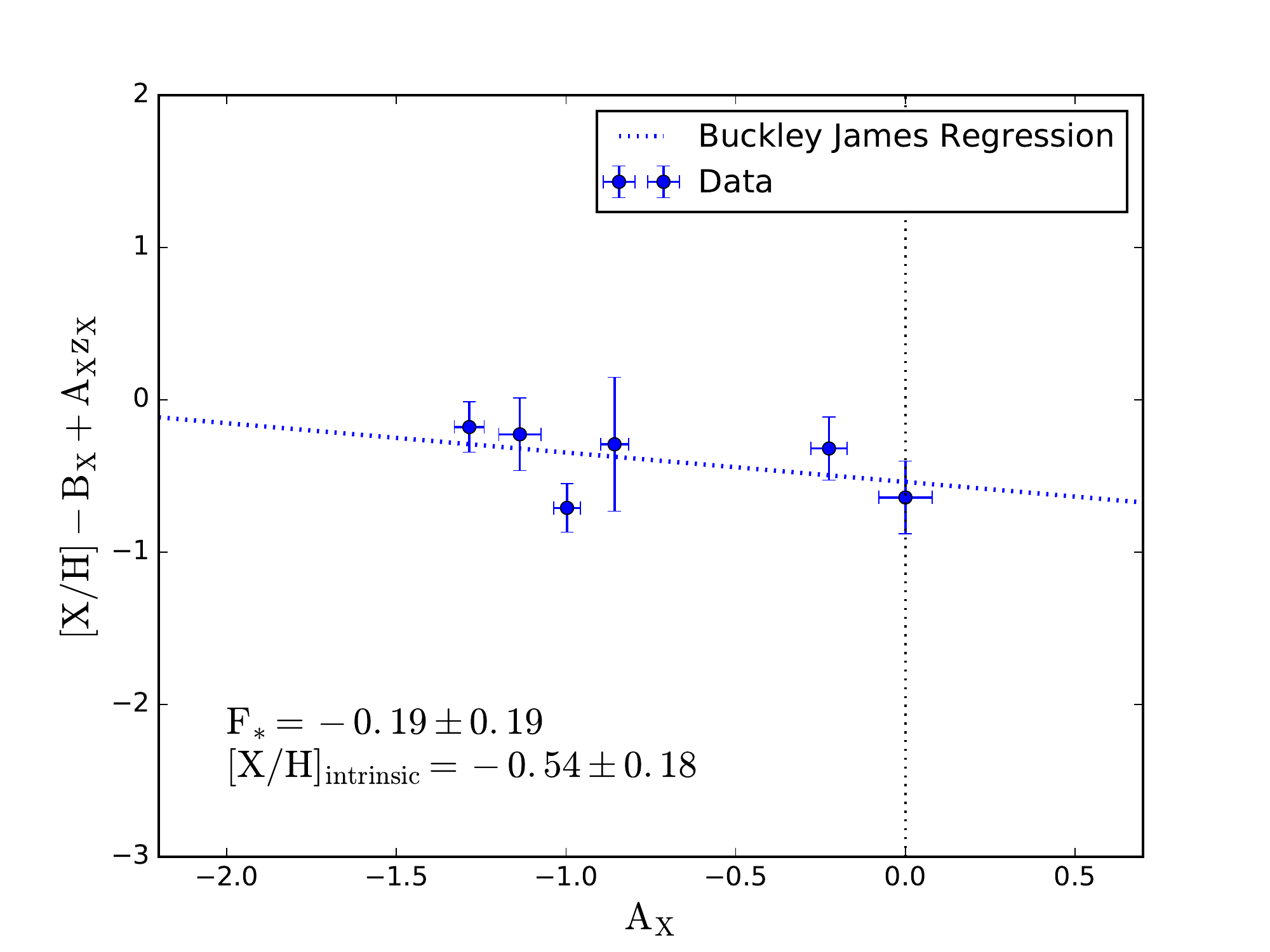}
\caption{{\bf Multi-element analysis of the metallicity of the absorber.} The method follows \citet{quiret16} and is based on the analysis developed by \citet{jenkins09} to estimate the dust content of the absorber. The method compares the dust depletion of dense neutral hydrogen systems to that of the Interstellar Medium (ISM) of our Galaxy. The fit is represented by two parameters: F$_*$ which is the line-of-sight depletion factor and A$_x$ which is the propensity of the element X to increase the absolute value of its particular depletion level as F$_*$ becomes larger. We find a F$_*$ value of $-$0.19$\pm$0.19 leading to a dust-free metallicity [X/H]=$-$0.54$\pm$0.18 (value at A$_x$=0), corresponding to 12+log(O/H)=8.15$\pm$0.20. This F$_*$ value indicates a galaxy with low dust content, corresponding to a log A$_V \sim -$1.70 \citep{vladilo06}.}
\label{f:Jenkins}
\end{figure}

\begin{table*}
\begin{center}
\caption{{\bf Absorption Properties of the sub-DLA towards Q2131$-$1207.}  The parameter $f_{\rm H2}$ = 2N(H2)/[N(\hi)+2N(H2)] indicates the molecular fraction in the absorber.
The ionisation parameter U is derived from the  photo-ionisation modelling. Here, we apply the multi-element analysis proposed by \citet{jenkins09} to estimate the dust content of the absorber \citep{quiret16}. The derived value F$_*$ indicates a galaxy with low dust content. The resulting metallicities and extinction are provided in the last columns of the table. The quoted errors are 1$\sigma$ uncertainties.}
\label{t:AbsProp}
\begin{tabular}{lcccccccc}
\hline\hline
Quasar Field		 &$z_{\rm abs}$ &$\log N(H\,\textsc{i})$			&log $f_{\rm H2}$  &log U	&F$_*$  &[X/H]$_{\rm  dust-free}$ &12+log(O/H)   &log A$_V$\\

&&[cm$^{-2}$]&&&&&\\
\hline
Q2131$-$1207   	&0.42980		  &19.50$\pm$0.15$^1$	  &$-$2.84$\pm$0.17$^1$  
&$-$5.6$^1$ &$-$0.19$\pm$0.19$^2$  &$-$0.54$\pm$0.18$^2$ &8.15$\pm$0.20$^2$ &$-$1.70$^2$\\
\hline\hline 				       			 	 
\end{tabular}			       			 	 
\end{center}			       			 	 
\begin{minipage}{180mm}
{\bf References:} {\bf $^1$:} \citet{muzahid16} {\bf $^2$:} This work\\
\end{minipage}
\end{table*}

The absorption system at z$_{\rm abs}$=0.42980 in the spectrum of the quasar Q2131$-$1207 was originally reported by \citet{weymann79} as a \mgii\ and \feii\ absorber. \citet{rao06} measured $\log [N$(H\,\textsc{i})/cm$^{-2}$]\,$= 19.18 \pm 0.03$ from archival HST/FOS spectra. \citet{muzahid16} used recent HST/COS FUV data to derive $\log [N$(H\,\textsc{i})/cm$^{-2}$]\,$= 19.50 \pm 0.15$ which we adopt in the following.

\citet{som15} used HST/COS NUV spectroscopy to determine the absorption metallicity (S, Si and C) of the absorber using a 5-component velocity structure over $\sim 100$ km s$^{-1}$. They further perform detailed photoionisation modelling based on the CLOUDY software to estimate the ionisation fraction of gas in this sub-DLA. Using the \siiii/\siii\ ratio, they deduce an ionisation parameter log U$<-$2.8. The additional HST/COS FUV quasar spectra presented by \citet{muzahid16} confirm these results while providing additional information on the absorption properties of the system. Abundances of O, S, Si, C, N, Ar, Fe, Mg, Mn and Ca have been obtained from HST/COS for the first six elements and Keck/HIRES archival spectra for the remaining \citep{som15, muzahid16}. Excerpts from the velocity profiles in \feii, \mgii\ and \mgi\ are shown in Figure~\ref{f:HIRES}. The zero velocity component is set to the redshift of galaxy "a", z$_{\rm gal}$=0.43005. Most of the absorption appears to lie bluewards of this systemic redshift. We note however one weak component at $\sim +50$ km~s$^{-1}$ which is only seen in the strongest transitions (i.e. \mgii). The CLOUDY modelling by \cite{muzahid16} lead to a density of $\log(n_{\rm H}/{\rm cm}^{-2}) \sim -0.2$ corresponding to a density $n_{\rm H} \sim 0.6$\,cm$^{-3}$ and an ionisation parameter of log U$\sim -$5.6 in agreement with earlier claims. The resulting ionisation corrections vary from element to element, but are, as expected, negligible for \oi, leading to an absorbing metallicity with respect to solar of [O/H]=$-$0.26$\pm$0.19 (1$\sigma$ uncertainty). 

Here, we apply the multi-element analysis proposed by \citet{jenkins09} to estimate the dust content of the absorber. We refer the reader to \citet{quiret16} for a complete description of the method applied to quasar absorbers. In short, the method compares the dust depletion of dense neutral hydrogen systems to the gas-phase element abundances reported in the literature for 17 different elements sampled over 243 sight lines in the local part of our Galaxy. The trend of the depletions into solid form (dust grains) as a function of elements is characterised by two parameters: F$_*$ which is the line-of-sight depletion factor and A$_x$ which is the propensity of the element X to increase the absolute value of its particular depletion level as F$_*$ becomes larger. We perform 100,000 realisations of the Buckley-James linear regression which treats limits using a survival analysis technique. In this case, we choose to exclude the \ci\ upper limit because it appears to be inconsistent with the other measures. The determination of \oi\ (A$_x$=$-$0.225$\pm$0.053) and \ni\ (A$_x$=$-$0.000$\pm$0.079) column densities which have A$_x$ values near zero means that the leverage on A$_x$ is good as illustrated in Fig~\ref{f:Jenkins}. We find a F$_*$ value of $-$0.19$\pm$0.19 leading to a dust-free metallicity [X/H]=$-$0.54$\pm$0.18 (value at A$_x$=0), corresponding to 12+log(O/H)=8.15$\pm$0.20 (using the solar value 12+log(O/H)=8.69 from \citet{asplund09}. This F$_*$ value indicates a galaxy with low dust content, corresponding to an extinction log A$_V \sim -$1.70 \citep{vladilo06}. 

 In addition, \citet{muzahid16} report the detection of molecular hydrogen in the absorber in one component at z$_{\rm abs}$=0.42981. This is at odds with findings from \citet{ledoux03} who report that absorbers with H$_{\rm 2}$ detections are usually amongst those having the largest depletion factors. Their fit results in a large total molecular fraction, log $f_{\rm H2}$=$-$2.84$\pm$0.17. However, no absorption lines from J$>$3 levels are observed possibly due to the absence of a local UV radiation field. A single excitation temperature can explain all of the remaining level populations, T$_{\rm ex}$=206$\pm$6 K, indicating the presence of cold gas. The authors note that the physical conditions in this system are similar to those of diffuse molecular clouds in the Galactic halo \citep{jenkins73, spitzer95}. The absorption properties of the sub-DLA are summarised in Table~\ref{t:AbsProp}.

\begin{figure}
\includegraphics[angle=0,scale=0.23]{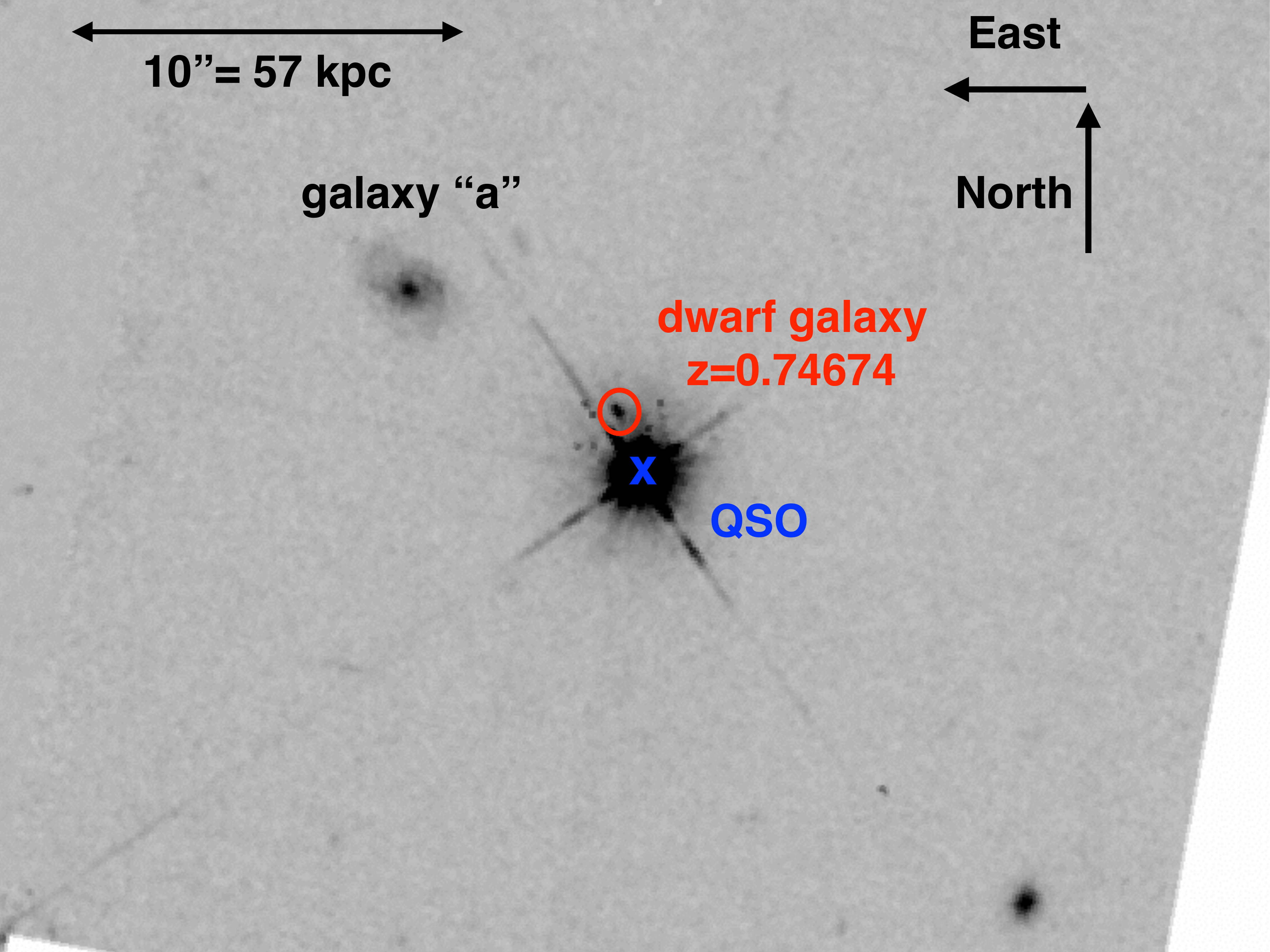}
\caption{{\bf HST/WFPC2 F702 image of the quasar field.} The  quasar position is marked with a blue cross. Galaxy "a" is clearly detected. The red circle shows the position of the dwarf galaxy at impact parameter $\delta$=2.06" away from the quasar line-of-sight. The redshift is z$_{\rm gal}$ =0.74674, i.e. higher than the redshift of the quasar. The MUSE spectrum of that object is shown in Figure~\ref{f:NearObj} of the Appendix. }
\label{f:hst}
\end{figure}

\subsection{Imaging: a Large Galaxy}

Bergeron (1986) discovered a galaxy located 8.6$\arcsec$ away from the quasar from ground-based broad-band imaging. A spectrum confirms that the object is at the redshift of the absorber thanks to the detection of the \oii\ doublet, H$\beta$, Ca H\&K absorption doublet and a prominent break at the Balmer limit (see also \citet{guillemin97}. This was the first spectroscopic confirmation of a galaxy associated with a quasar absorber. \citet{kacprzak11b} used HST/WFPC2 F702W archival data shown in Figure~\ref{f:hst} and reported a B-band absolute magnitude $M_{\rm B} = -20.32$ and a corresponding luminosity of $0.46 L_\ast$. They further studied the morphological properties of the object using GIM2D modelling \citep{simard02} and derived an inclination angle sin i=0.75$\pm$0.05 and half light radius r$_{1/2}$=4.65 kpc. \citet{muzahid16} used a Keck/ESI flux uncalibrated spectrum to report detections of H$\alpha$, \nii\ and \oiii\ emission lines. They used the first two lines to estimate an emission metallicity of 12+log(O/H)=8.68$\pm$0.09 from the N2 index \citep{pettini04}. They also argued, based on simple modelling, that the profiles of the absorption lines in the sub-DLA are inconsistent with those expected from a rotating disk associated with this galaxy, but the conclusion is heavily based on a mismatch in velocity space between the absorption profile and the rotation curve of the galaxy for which they derive an emission redshift z$_{\rm gal}$=0.42966.

\citet{muzahid16} note a dwarf galaxy nearer to the line-of-sight of the quasar (at impact parameter $\delta = 2.06^{\prime\prime}$, circled in red in Figure~\ref{f:hst}) in the HST images and speculate it may be the source of H$_2$ absorption, although no spectroscopy was available at the time to confirm the redshift of that object. The object is also clearly detected in the MUSE data presented here and shows a number of strong emission lines including \oii, \neiii, H$\delta$, H$\beta$, \oiii\ at z$_{\rm gal}$=0.74674, i.e. higher than the redshift of the quasar. A portion our the MUSE spectrum of that object is reproduced in Fig~\ref{f:NearObj} of the Appendix. This chance alignment once more calls for caution when associating absorbers with galaxies without spectroscopic information \citep{peroux11a}.

\subsection{New MUSE Observations}
Although the absorbing galaxy was known prior to our observations, the MUSE field was centered on the quasar to maximise environmental studies. The observations were carried out in service mode (under programme ESO 96.A-0303 A) at the European Southern Observatory on the 8.2 m YEPUN telescope. The seeing constraint for these observations was $<0.8\arcsec$ and natural seeing mode was used. Two "Observing Blocks" (OBs) were taken on the nights of 4$^{th}$ and 8$^{th}$ of October 2015, respectively. The field was rotated by 90 degrees between the two OBs. Each of these observations are 1200 sec long. These were further divided into two equal sub-exposures, with an additional field rotation of 90 degrees and sub-arcsec dithering offset in 2-step pattern to minimise residuals from the slice pattern. The field of view is 59.9 arcsec $\times$ 60 arcsec, corresponding to a 0.2 arcsec/pixel scale. We used the "nominal mode" resulting in a spectral coverage of $\sim$4800-9300 \AA. At the redshift of the target (z$\sim$0.4), the data cover emission lines from \oii~$\lambda \lambda 
3727, 3729$ to \oiii~$\lambda 5008$. The spectral resolution is R=1770 at 4800 \AA\ and R=3590 at 9300 \AA\ resampled to a spectral sampling of 1.25 \AA/pixel. A journal of observations summarising the properties of the targeted quasar is presented in Table~\ref{t:JoO}. 

\begin{table*}
\begin{center}
\caption{{\bf Journal of MUSE Observation of Q2131$-$1207.}  }
\begin{tabular}{lccccccccc}
\hline\hline
Quasar Field		 &Alternative Name &RA &DEC  &$z_{\rm QSO}$  &B mag &V mag &R mag & T$_{\rm exp}$  [sec] &seeing ["]\\
\hline
Q2131$-$1207   &PHL 1598	&21 31 35.26	&-12 07 04.8	&0.501	&16.33	&16.11	&15.87	&2$\times$1200 & 0.72$\pm$0.02 \\
\hline\hline 				       			 	 
\label{t:JoO}
\end{tabular}			       			 	 
\end{center}			       			 	 
\end{table*}

The data were reduced with version v1.6 of the ESO MUSE pipeline \citep{weilbacher15} and additional external routines for 
sky subtraction and extraction of the 1D spectra. Master bias, flat field images and arc lamp exposures based on data taken closest in time to the science frames were used to correct each raw cube. We checked that the flat-fields are the closest possible to the science observations in terms of ambient temperature to minimise spatial shifts. In all cases, we found the temperature difference to be less than 0.27 degrees, below the canonical 0.5 degrees set to be the acceptable limit. Bias and flat-field correction are part of the ESO pipeline. The raw science data were then processed with the $scibasic$ and $scipost$ recipes. During this step, the wavelength calibration was corrected to a heliocentric reference. We checked the wavelength solution using the known wavelengths of the night-sky OH lines and found it to be accurate within 25 km~s$^{-1}$. The individual exposures were registered using the point sources in the field within the $exp\_align$ recipe, ensuring accurate relative astrometry. Finally, the individual exposures were combined into a single data cube using the $exp\_combine$ recipe. The seeing of the final combined data is measured from the quasar and other bright point source in the data cube. The resulting point spread function (PSF) has a full width at half maximum of $0.72 \pm 0.02$ arcsec at 7000 \AA. 

The removal of OH emission lines from the night sky is accomplished with additional purpose-developed codes using two different methods. The $scipost$ recipe is first performed with sky-removal method turned off. The sky subtraction is then done with the ZAP (Zurich Atmosphere Purge) code \citep{soto16}. After masking bright objects in the field, ZAP uses a principal component analysis (PCA) to isolate the residual sky subtraction features and remove them from the observed datacube. In a second pass, the $scipost$ recipe is run with the sky subtraction method "simple" on, which directly subtracts a sky spectrum created from the data, without regard to the line spread function (LSF) variations. Again, after selecting sky regions in the field, we create PCA components from the spectra which are further applied to the science datacube to remove sky line residuals \citep{husemann16}. This second method is found to significantly improve the sky subtraction over large parts of the MUSE field-of-view and is chosen to process the data presented here.

The resulting flux calibration is compared with the R magnitudes of three known objects in the field, including the quasar itself, in order to estimate the flux uncertainties. The MUSE spectra of these objects are extracted and their broad-band fluxes are computed using an R filter passband. These fluxes are computed as AB magnitudes directly using the following relation:

\begin{equation}
 AB = -2.5\log_{10}(F_{\lambda}) - 5\log_{10}(<\lambda>) - 2.406 
\label{eq:abmag}
\end{equation}

where $F$ is the flux in erg/s/cm$^2$/\AA\ and $<\lambda>$ the filter central wavelength in \AA. The differences are found to be small (0.07 mag) and the mean of the differences of three reference objects provide an estimate on the flux error of $\pm 12$\%.

\section{Analysis}
%%%%%%%%%%%%%%%%%%%%%%%%%%%%%%%%%%%%%%%%%%%%%
%%%%%%%%%%%%%%%%%%%%%%%%%%%%%%%%%%%%%%%%%%%%%

\subsection{Galaxy Detections}
%%%%%%%%%%%%%%%%%%%%%%%%%%%%%%%%%%%%%%%%%%%%%

\begin{figure}
\hspace{-0.8cm}
\includegraphics[angle=0,scale=0.30]{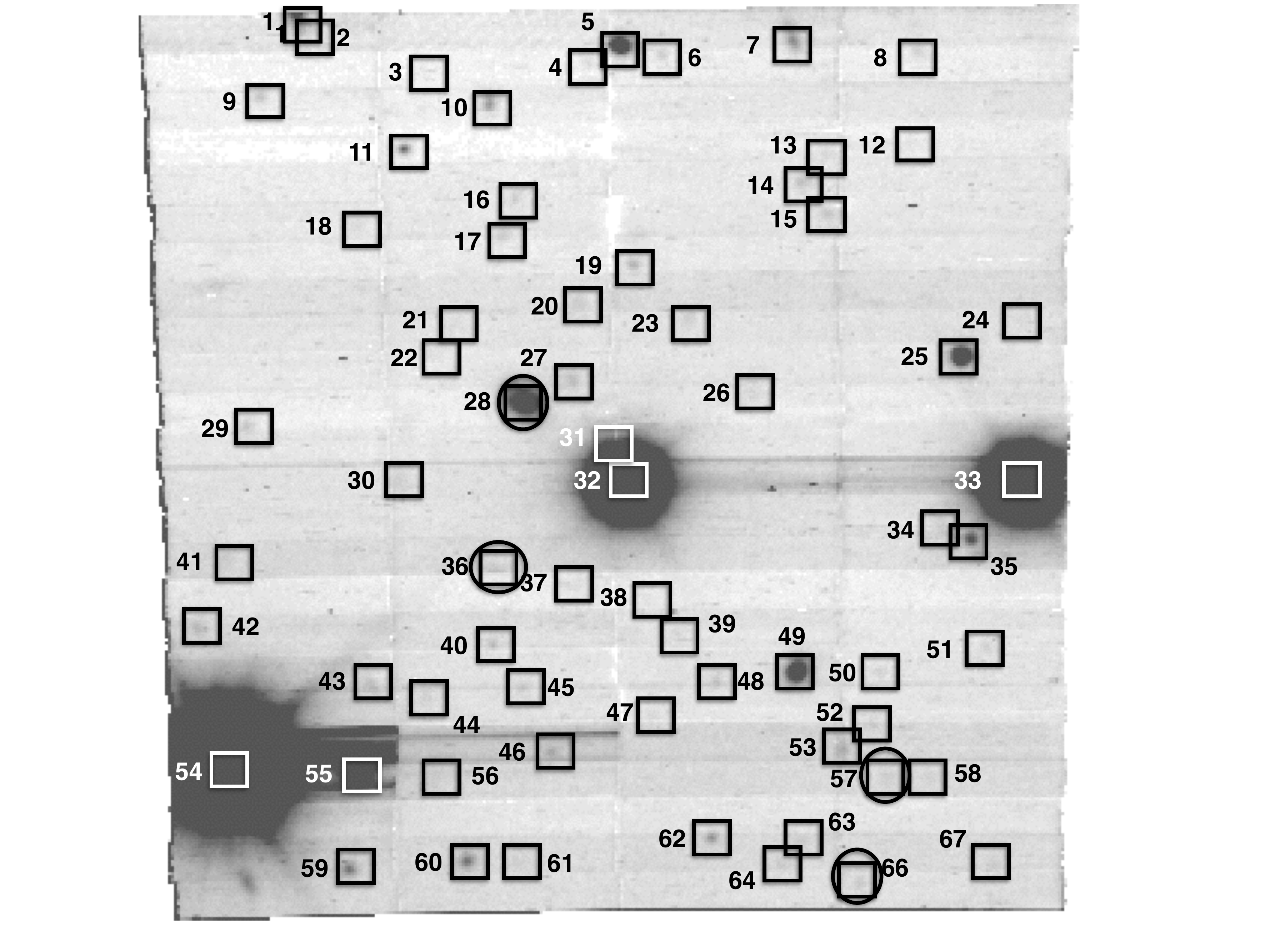}
\caption{{\bf Reconstructed white-light image of the combined exposures.} Orientation is north(up)-east(left) and the field-of-view is 59.9 arcsec $\times$ 60 arcsec. All the objects identified in the field are marked with a black or white (colour depending on contrast) square with an associated id number. The quasar is object 32, the nearby dwarf galaxy at $z_{\rm gal} = 0.74674$ is object 31 and the strong emitter at $z_{\rm gal} = 4.8986$, coincident with the redshifted wavelength of \oiii\,$\lambda 5008$ at the redshift of the sub-DLA is object 38. Objects 28, 36, 57 and 66 shown with an additional circle have redshifts consistent with the absorption redshift of the sub-DLA present in the quasar spectrum.}
\label{f:WhiteLight}
\end{figure}

\begin{figure}
\includegraphics[angle=0,scale=0.23]{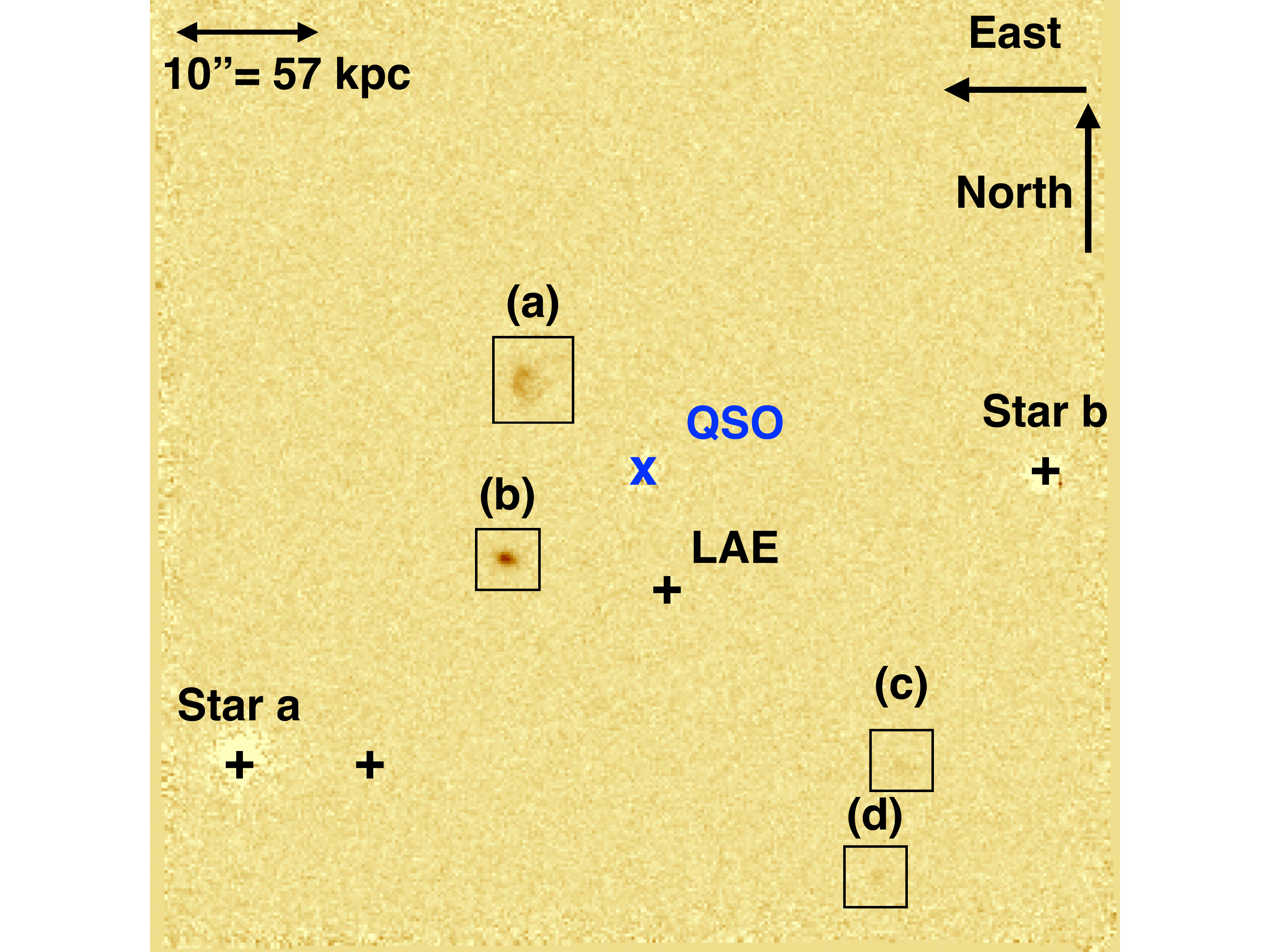}
\caption{{\bf A pseudo narrow-band filter around \oiii~$\lambda 5008$ at the redshift of the absorber.} The PSF of the bright objects in the field are removed and marked with black crosses but for the quasar which is shown as a blue cross. The spectrum of the object labelled \lya\ emitter (LAE) is shown in Figure~\ref{f:LyaObj} of the Appendix. The four emitters at the redshift of the sub-DLA are shown with squares and labelled "a" to "d". These are the same objects as labelled 28, 36, 57, and 66 in Figure~\ref{f:WhiteLight}.}
\label{f:NB}
\end{figure}

\begin{figure}
\includegraphics[angle=0,scale=0.23]{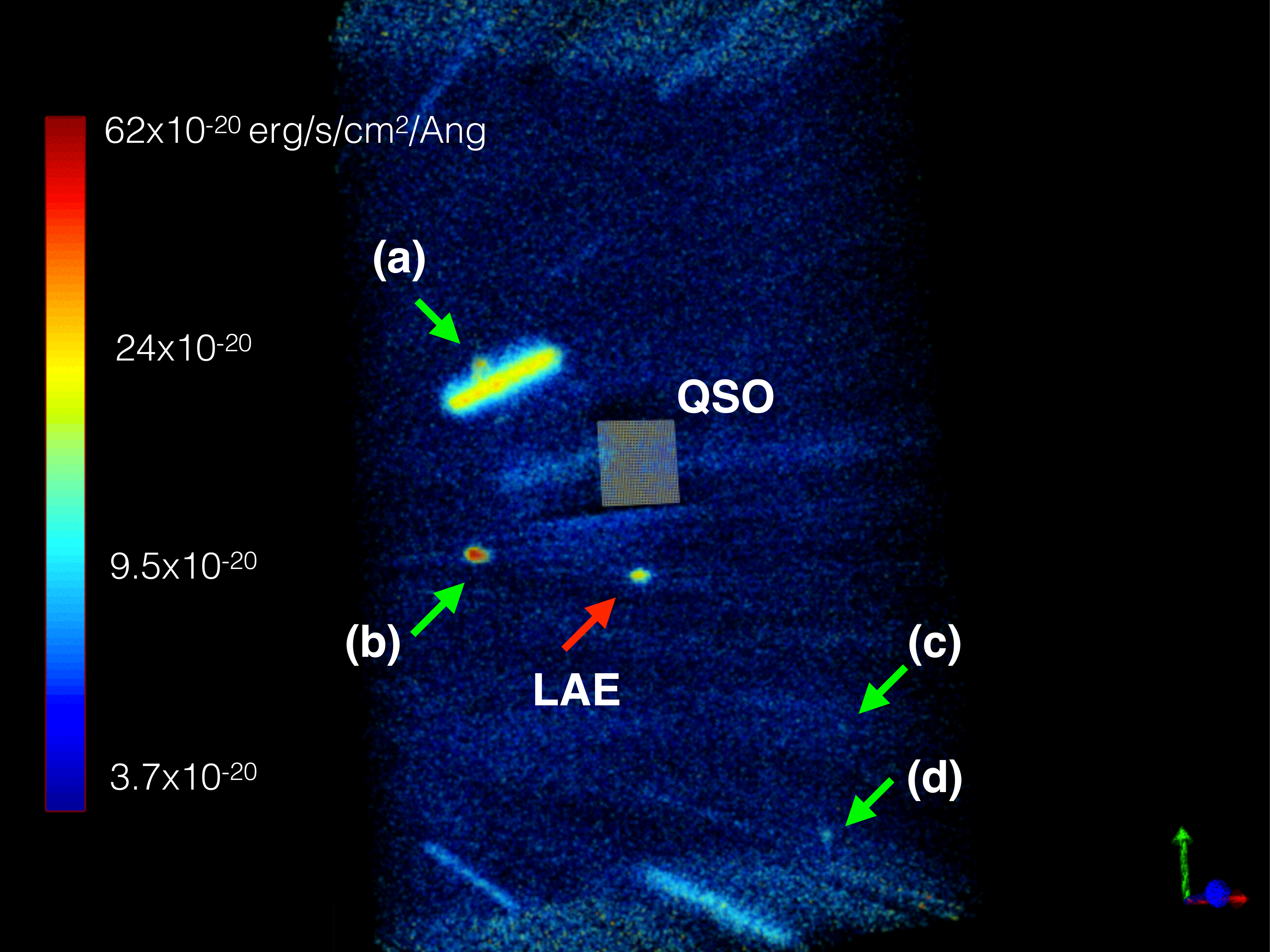}
\caption{{\bf A 3D rendering of the pseudo narrow-band filter around \oiii~$\lambda 5008$ at the redshift of the absorber.} The colour bar indicates fluxes in units of $10^{-20}$ erg/s/cm$^2$/\AA. Orientation is north(up)-east(left). Continuum-detected objects are shown as cylinders. The quasar has been subtracted and its location is shown as a grid in the wavelength plane. The emitters at the redshift of the sub-DLA are indicated with green arrows. The northern most object is clearly detected in the continuum while others are not. The red arrow indicates a background \lya\ emitter at z$_{\rm gal}$ =4.8986, whose spectrum is shown in Figure~\ref{f:LyaObj} of the Appendix. }
\label{f:3Drendering}
\end{figure}

Fig~\ref{f:WhiteLight} presents the reconstructed white-light image of the combined exposures. All the objects identified in the field are marked with a black or white square with an associated id number. Fig~\ref{f:NB} presents a pseudo narrow-band filter around \oiii~$\lambda 5008$ at the redshift of the absorber. The PSFs of the bright objects in the field are removed and marked with black crosses but for the quasar which is shown as a blue cross. Fig~\ref{f:3Drendering} is a 3D-rendering of the same narrow wavelength slice of the cube. One object appears to be coincident with the redshifted wavelength of \oiii\,$\lambda 5008$ at the redshift of the sub-DLA. However, the emission line is undoubtedly Ly$\alpha$ from a galaxy at $z_{\rm gal} = 4.8986$, as indicated by  the characteristic asymmetric emission + absorption line profile and the presence of narrow N\,\textsc{v}\,$\lambda 1240.81$ emission (Figure~\ref{f:LyaObj} of the Appendix). This object is marked as "LAE" in Figure~\ref{f:NB} and shown with a red arrow in Figure~\ref{f:3Drendering}.

In these MUSE data, we recover the known large galaxy (galaxy "a") reported by \citet{bergeron86} at an angular distance of 9.2" from the quasar. This bright object is clearly extended and displays spiral arms or possibly tidal tails from an interacting system. In addition to this object, we report 3 new detections at the redshift of the sub-DLA. One of these, dubbed galaxy "b" is at a similar angular distance from the quasar as galaxy "a", $10.7 ^{\prime\prime}$. It is significantly fainter than galaxy "a" and barely visible in the HST images \citep{muzahid16}. Two other objects, galaxies "c" and "d" are further away at 26.0" and 30.7" respectively, corresponding to 147 and 174 kpc at the redshift of the sub-DLA and fall outside the HST field. While we consider these two galaxies to be too far away from the quasar sightline to be the hosts of the sub-DLAs \citep{peroux03}, they are likely to trace a more general structure at that redshift. The sky location of each of these objects is reported in Table~\ref{t:GalDetect}, including a reference id from Fig~\ref{f:WhiteLight}.

\begin{figure*}
\includegraphics[angle=0,width=13.cm,height=8.cm]{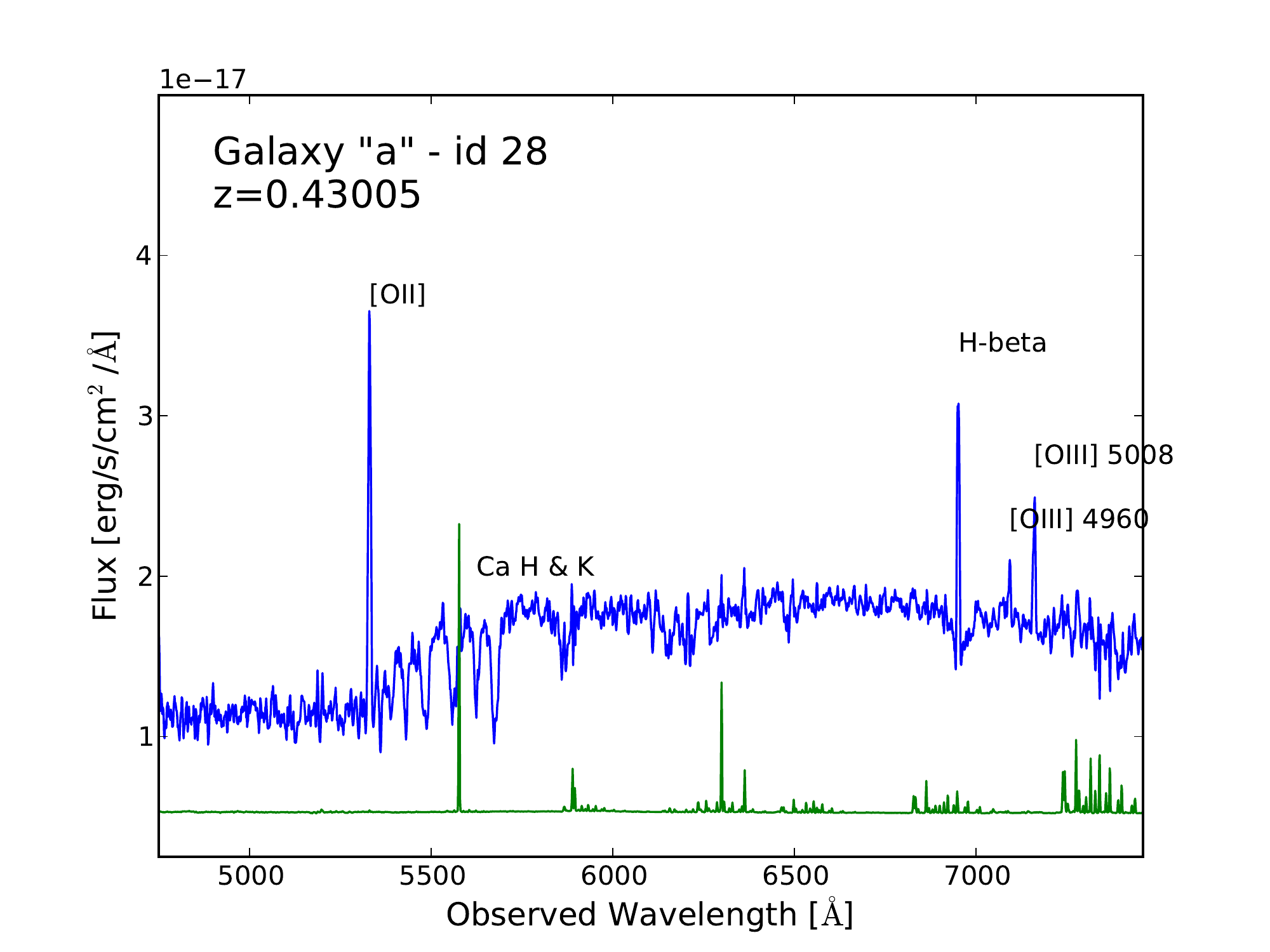}
\caption{{\bf MUSE spectrum of galaxy "a".} In this and the following figure, the MUSE spectrum is shown in blue and the sky spectrum is shown in green (not to scale and arbitrarily offset). The flux scales are indicated above the top-left corner of the panel. The object lies $\delta$=9.2" (52 kpc) away from the quasar line-of-sight at z$_{\rm gal}$=0.43005. The spectrum is smoothed (3-pixel boxcar) for display purpose. The continuum is clearly detected, showing also numerous emission lines of the unresolved \oii\ doublet, \hd, \hb\ and \oiii\ as well as absorption lines from Ca H\&K and a prominent Balmer decrement. Some other absorption lines are due to tellurics.}
\label{f:Gal_a}
\end{figure*}

\begin{figure*}
\hspace{-0.8cm}
\includegraphics[angle=0,height=8.cm,width=7.5cm]{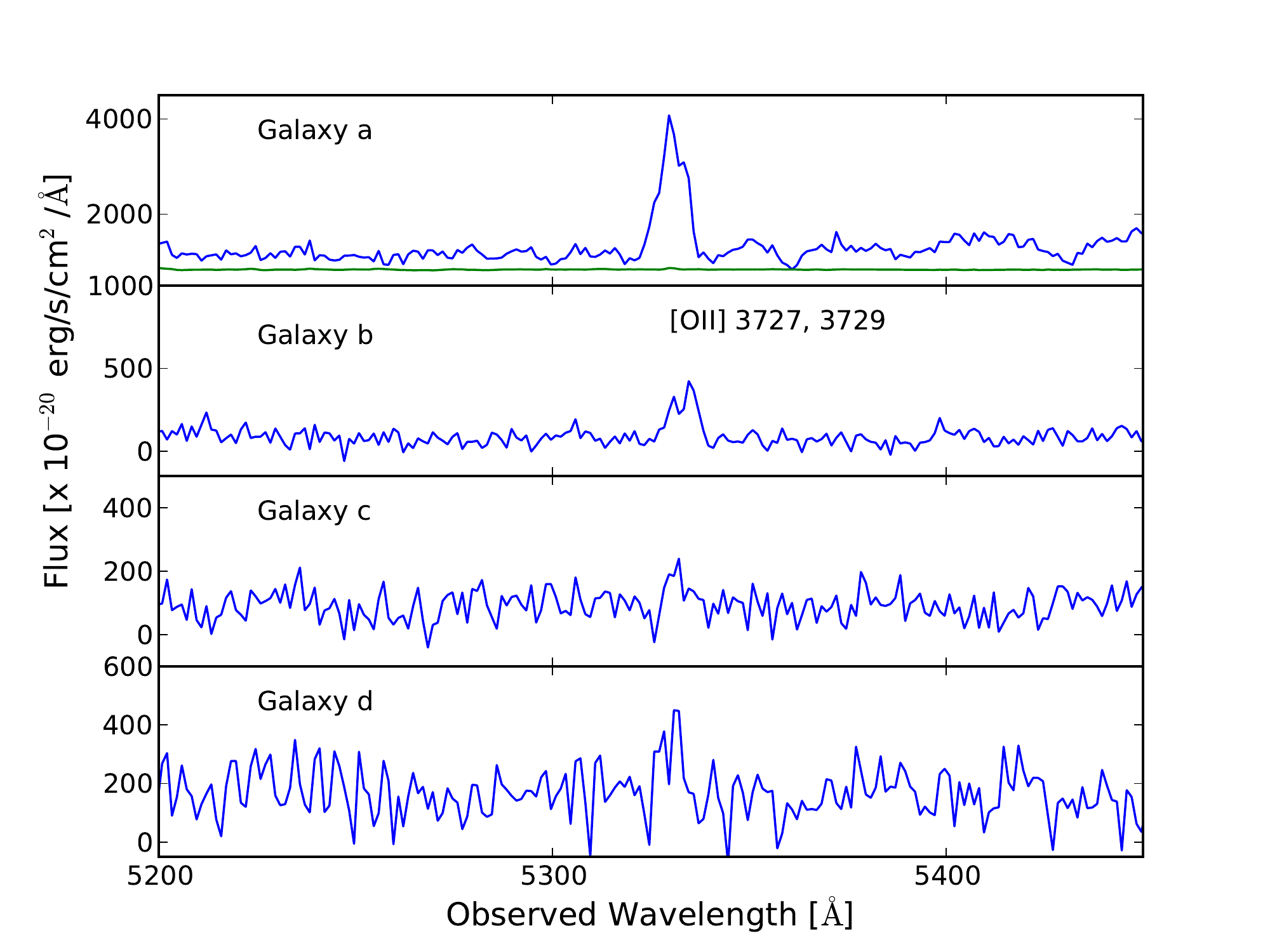}
\hspace{-0.8cm}
\includegraphics[angle=0,height=8.cm,width=11.5cm]{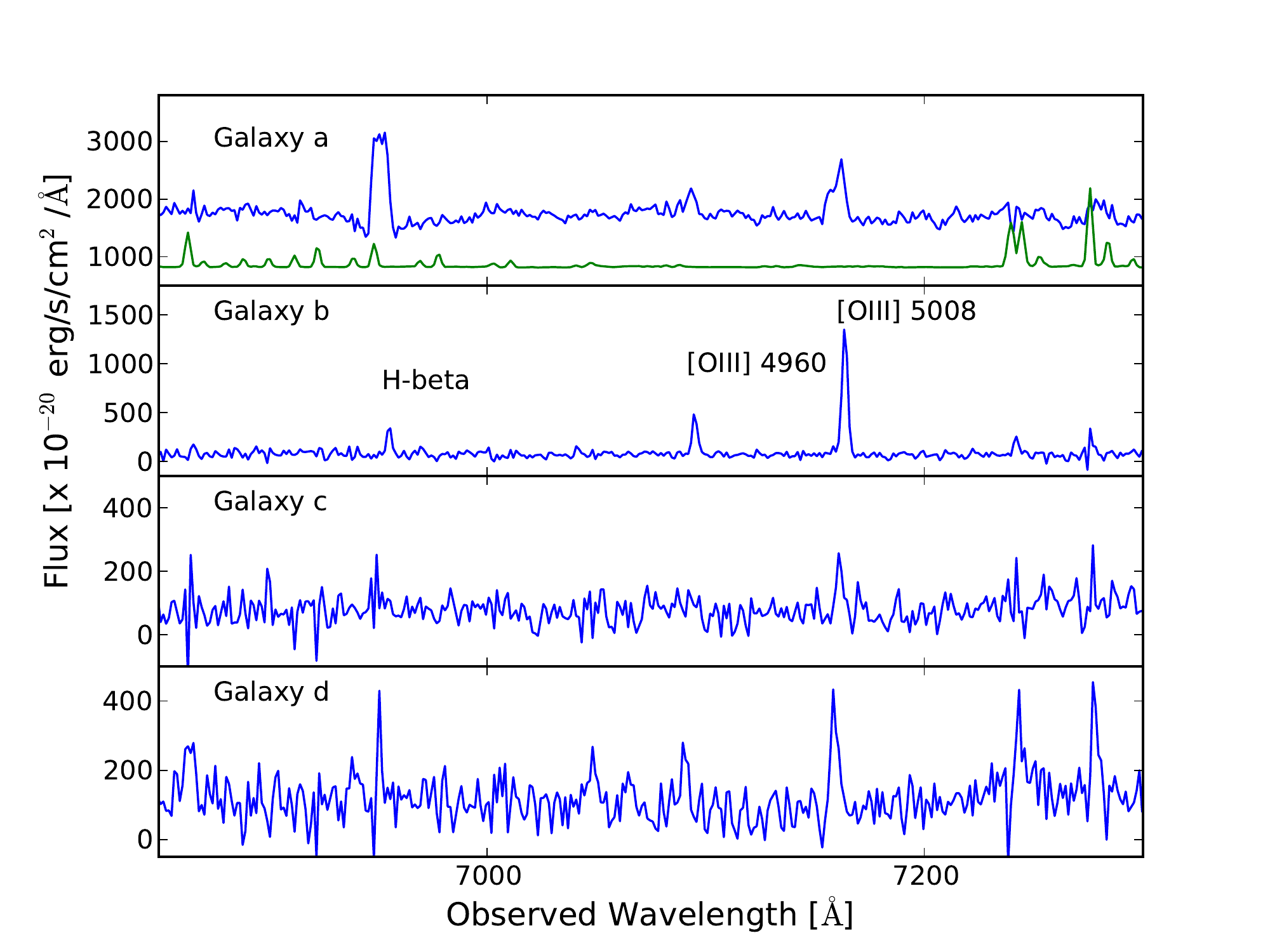}
\caption{{\bf MUSE spectra of galaxies "a" to "d"}. The \oii\ (left panel) and \hb\ and \oiii\ (right panel) regions of all four galaxies are shown. The four objects have redshifts consistent with that of the quasar absorber.}
\label{f:All_Gal}
\end{figure*}

\begin{table*}
\begin{center}
\caption{{\bf Sky location of the four objects at the redshift of the absorber.} Sky offsets are relative to the central position of the quasar. $\delta$ is the angular distance in arcsec and the impact parameter in kpc. The redshift of the strongest component of the absorption profile is z$_{\rm abs}$=0.42980. The broad-band R magnitudes and luminosity relative to L$*$ are also provided.
}
\begin{tabular}{lccccccccc}
\hline\hline
Galaxy & id Fig~\ref{f:WhiteLight}		 &$\Delta$RA &$\Delta$Dec  &$\delta$ &$\delta$ &$z_{\rm gal}$ &M$_{\rm R}$&L/L$*$ &Reference \\
	&	 &["] &["]  &["] &kpc &&& &\\
\hline
Galaxy a &28   &$+$7.2	&$+$5.7	&9.2		&52	&0.43005  &$-$21.16 &0.60 &Bergeron et al. 1986, this work \\		
Galaxy b &36   &$+$8.9	&$-$5.9	&10.7	&61	&0.43072  &$-$19.70 &0.15 &this work \\				
Galaxy c &57   &$-$17.2	&$-$19.5  &26.0	&147	&0.43006  &$-$16.38 &0.01 &this work \\				
Galaxy d &66   &$-$15.5	&$-$26.4 &30.7	 	&174	&0.42982	 &$-$16.25 &0.01 &this work \\				
\hline\hline 				       			 	 
\label{t:GalDetect}
\end{tabular}			       			 	 
\end{center}			       			 	 
\end{table*}

To extract 1D spectra for each of these galaxies, we adjust the radius according to the galaxy size to be (for galaxy "a" to "d"): 10, 5, 3 and 8 pixels respectively, and add the 1D spectra for each spaxel within these radii. Fig~\ref{f:Gal_a} presents the resulting whole spectrum for galaxy "a". The continuum of galaxy "a" is clearly detected, showing also numerous emission lines of the unresolved \oii\ doublet, \hd, \hb\ and \oiii\ as well as absorption lines from Ca H\&K and a prominent Balmer decrement. We note that the absorption lines from Ca H\&K aligned well with the systemic redshift of the galaxy so that he "down-the-barel" technique indicates no sign of gas flows at this spectral resolution. The remaining galaxies have a weak continuum but all show the five emission lines listed above, except for galaxy "c" for which an upper limit in \oiii\ 4960 \AA\ can be derived and galaxy "d" for which an upper limit in \oiii\ 5008 \AA\ can be derived. We derive a 3$\sigma$ upper limit for non-detection based on the measured rms at the expected position of the line for an unresolved source spread over 6 spatial pixels and spectral FWHM = 2.4 pixels = 3 \AA. Figure~\ref{f:All_Gal} displays the \oii\ (left panel) and \hb\ and \oiii\ (right panel) regions of all four galaxies.

\begin{table*}
\begin{center}
\caption{{\bf Physical properties of the detected galaxies.} The fluxes are expressed in erg/s/cm$^2$. The redshifts are determined from the 1D MUSE spectra calibrated in air. The SFR is estimated from \oii\ fluxes without dust correction. The metallicity is based on the R$_{\rm 23}$  index. For comparison, the metallicity of the neutral gas probed in absorption is 12+log(O/H) = 8.15$\pm$0.20. The quoted errors are 1$\sigma$ uncertainties.
}
\begin{tabular}{lcccccc}
\hline\hline
Galaxy		 &F(\oii)  &F(\hb)   &F(\oiii) 4960   &F(\oiii)  5008	&SFR  	&12+log(O/H) \\
		 &[erg/s/cm$^2$] &[erg/s/cm$^2$] &[erg/s/cm$^2$]  &[erg/s/cm$^2$]&[$M_{\odot}$/yr]& \\
\hline
Galaxy a  &21.4$\pm$2.1 $\times$ 10$^{-17}$ &16.3$\pm$1.6 $\times$ 10$^{-17}$ &2.2$\pm$0.2 $\times$ 10$^{-17}$ &6.7$\pm$0.7 $\times$ 10$^{-17}$  &2.0$\pm$0.2	&8.98$\pm$0.02 \\	
Galaxy b  &2.4$\pm$0.2 $\times$ 10$^{-17}$  &0.8$\pm$0.1 $\times$ 10$^{-17}$  &1.2$\pm$0.2 $\times$ 10$^{-17}$ &4.2$\pm$0.4 $\times$ 10$^{-17}$  &0.2$\pm$0.1	&8.32$\pm$0.16\\	
Galaxy c  &0.8$\pm$0.1 $\times$ 10$^{-17}$  &0.2$\pm$0.1 $\times$ 10$^{-17}$  &$<$0.6 $\times$ 10$^{-17}$       &0.6$\pm$0.1 $\times$ 10$^{-17}$  &0.08$\pm$0.1  &-- \\	
Galaxy d  &1.5$\pm$0.1 $\times$ 10$^{-17}$  &0.5$\pm$0.1 $\times$ 10$^{-17}$  &0.7$\pm$0.1 $\times$ 10$^{-17}$ &$<$1.2 $\times$ 10$^{-17}$  &0.1$\pm$0.1	&-- \\	
\hline\hline 				       			 	 
\label{t:GalFlux}
\end{tabular}			       			 	 
\end{center}			       			 	 
\begin{minipage}{180mm}
Note: the 3$\sigma$ upper limit for non-detection is computed for an unresolved source spread over 6 spatial pixels and spectral FWHM = 2.4 pixels = 3 \AA.  \\
\end{minipage}
\end{table*}			       			 	 

The fluxes of the detected emission lines are estimated from Gaussian fit.  The \hb\ emission line in galaxy "a" has the characteristic broad stellar absorption line component at the base of the emission line which we have removed before fitting the emission line itself \citep{zych07}. We do not correct the emission lines fluxes for reddening because \ha\ is not covered by our data and and \hg\ is too weak. In addition, we use the fit of four of the emission lines (namely the \oii\ doublet, \hb, \oiii\ $\lambda 4960$ and \oiii\ $\lambda 5008$) to estimate the redshifts of the emitting galaxies. For galaxy "a", we derive z$_{\rm gal}$=0.43005 also including \hd\ from the MUSE spectrum calibrated in air.  This value of z$_{\rm gal}$ is at odds with the earlier measure by \citet{muzahid16} who reported z$_{\rm gal}$=0.42966. We note that the dispersion between the measurements of the different lines is $\Delta$ z=0.00033, corresponding to $\Delta v = 70$\,km~s$^{-1}$. The physical properties of the four galaxies deduced from the emission line fits are collected in Table~\ref{t:GalDetect}. We find that the maximum redshift difference between these objects, $\Delta$ z=0.00090, which translates into a velocity difference of only $\Delta v \simeq 200$\,km~s$^{-1}$, significantly above our wavelength calibration accuracy (25\,km~s$^{-1}$). Finally, we build a MUSE cube with the Cousins R filter band-pass to compute the broad-band R magnitude and luminosities relative to L$*$ \citep{lin96} as listed in Table~\ref{t:GalDetect}.

\subsection{Star Formation Rates}
%%%%%%%%%%%%%%%%%%%%%%%%%%%%%%%%%%%%%%%%%%%%%

Based on the \oii\ detections, we derive the Star Formation Rate (SFR) using the prescription by \citet{kennicutt98} :

\begin{equation}
SFR_{[O II]} = (1.4\pm0.4) \times 10^{-41} L([O II])
\end{equation}

We note that the \oii\ doublet at the redshift of the sub-DLA is slightly affected by a weak sky line which happens to fall right in between the two lines of the doublet. The resulting SFRs, uncorrected for dust depletion, are listed in Table~\ref{t:GalFlux}. Galaxies "a" and "b" with SFR$\geq 0.2 M_{\odot}$/yr are star-forming galaxies while galaxies "c" and "d" are passive \citep{rauch08}.

\subsection{Metallicity}
%%%%%%%%%%%%%%%%%%%%%%%%%%%%%%%%%%%%%%%%%%%%%
Emission lines are commonly used to estimate gas-phase metallicities in extragalactic \hii\ regions. We use the R$_{23}$ indicator which was first introduced by Pagel et al. (1979), and is widely used for measuring the oxygen abundances if the fluxes of \oiii\ and \oii\ are known:  

\begin{equation}
\log{R_{23}} = \log\Biggr({{[O~II] \lambda\lambda 3727,3729 + [O~III]
\lambda\lambda4960,5008}\over{H\beta}}\Biggr)
\end{equation}

The resulting metallicities are derived using the relation of \citet{mcgaugh91} which accounts for the level of ionisation via the O32 parameter  (see \citet{kobulnicky99} for the relevant equations). The resulting values are listed in Table~\ref{t:GalFlux}. For galaxy "a", we use the metallicity derived from the N2-index indicator of \citet{pettini04} to exclude the lower branch. We note that for that object, our estimate of the metallicity (12+log(O/H) = 8.98$\pm$0.02) is higher than the one reported by \citet{muzahid16}: 12+log(O/H) = 8.68$\pm$0.09; which can be explained by the fact that they use the N2-index which is known to saturate at solar metallicities \citep{pettini04}. Thus, the R$_{23}$ index is probably more reliable in this regime\footnote{although we note that since we have not applied any dust extinction correction to the measured emission line fluxes, our value of 12+log(O/H) for galaxy "a" may be an overestimate.}. We also note that both these values are higher than the metallicity of the neutral gas probed in absorption reported in the earlier section: 12+log(O/H) = 8.15$\pm$0.20 (see Table~\ref{t:AbsProp}). For galaxy "b", we derive 12+log(O/H) = 8.32$\pm$0.16, i.e. a value consistent with the absorption metallicity within the errors. We cannot put stringent constraints on the metallicity estimates of galaxy "c" and "d".

In addition, for galaxy "a", we have been able to derive the metallicity gradient inside the galaxy based on two measures of the metallicity (centrally in a 3 pixel-radius and in annulus around it). We measure a slope +0.01$\pm$0.03 dex kpc$^{-1}$ consistent with a flat distribution of metallicities. We recall, that for the local galaxy gradients, the isolated spiral control sample of \citet{rupke10} indicates a median slope of $-$0.041$\pm$0.009 dex kpc$^{-1}$.These two values are consistent within the errors. Extrapolating the slope of +0.01$\pm$0.03 dex kpc$^{-1}$ to the impact parameter of galaxy "a" (52 kpc) would led to an expected metallicity 12+log(O/H) = 9.36$\pm$1.56 which is consistent with the one measured in absorption: 12+log(O/H) = 8.15$\pm$0.20. We warn the reader, however, that these two metallicity indicators provide a measure of the metallicity in different phases of the gas (\citet{peroux12, rahmani16}).

\subsection{Morphological and Kinematical Properties}
%%%%%%%%%%%%%%%%%%%%%%%%%%%%%%%%%%%%%%%%%%%%%
\begin{figure*}
\begin{center}
\includegraphics[width=14.cm, angle=0]{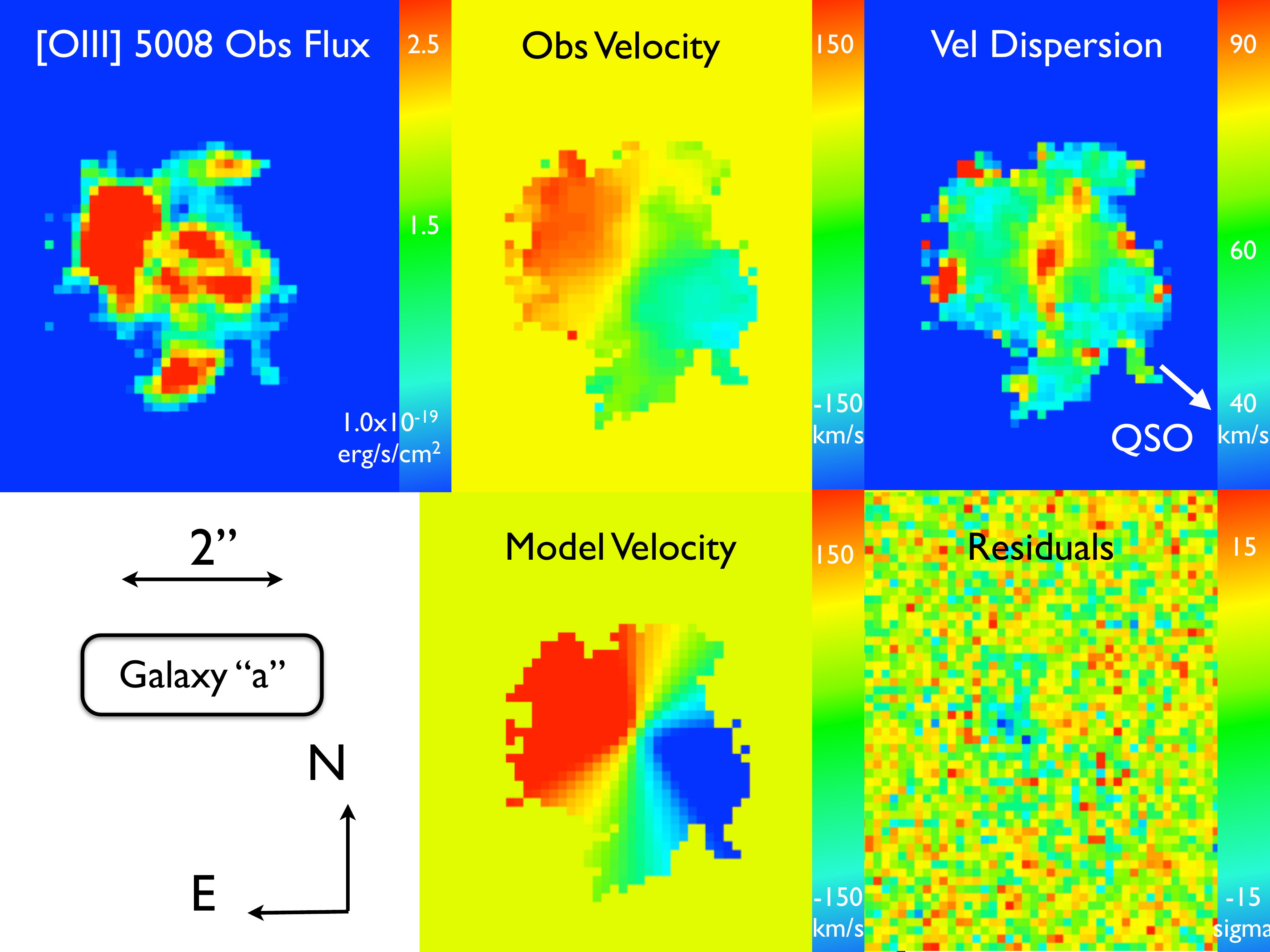}
\caption{{\bf Morpho-kinematic analysis of galaxy "a".} The maps in the top row show the MUSE observations. The maps in the bottom row illustrate the results from our 3D modelling. The bottom right panel shows the residuals from the subtraction of the model from the observations. The direction to the quasar is shown with a white arrow in the top-right panel. The orientation and scales are indicated on the figures. The maps clearly show the spiral arms of galaxy "a" or possibly tidal tails in an interacting system as well as sub-structure in \oiii\ on scales of 1". The velocity field and velocity dispersion (peaking in the center) maps clearly indicate rotation expected from an extended disk. We measure $V_{\rm max}$=200$\pm$3 km~s$^{-1}$ corresponding to $M_{\rm dyn}$=7.4$\pm$0.4$\times$10$^{10}$ M$_{\odot}$ and M$_{\rm halo}$=2.9$\pm$0.2$\times$10$^{12}$ M$_{\odot}$.
}
\label{f:Gal_a_Kine}
\end{center}
\end{figure*}

\begin{figure*}
\begin{center}
\includegraphics[width=14.cm, angle=0]{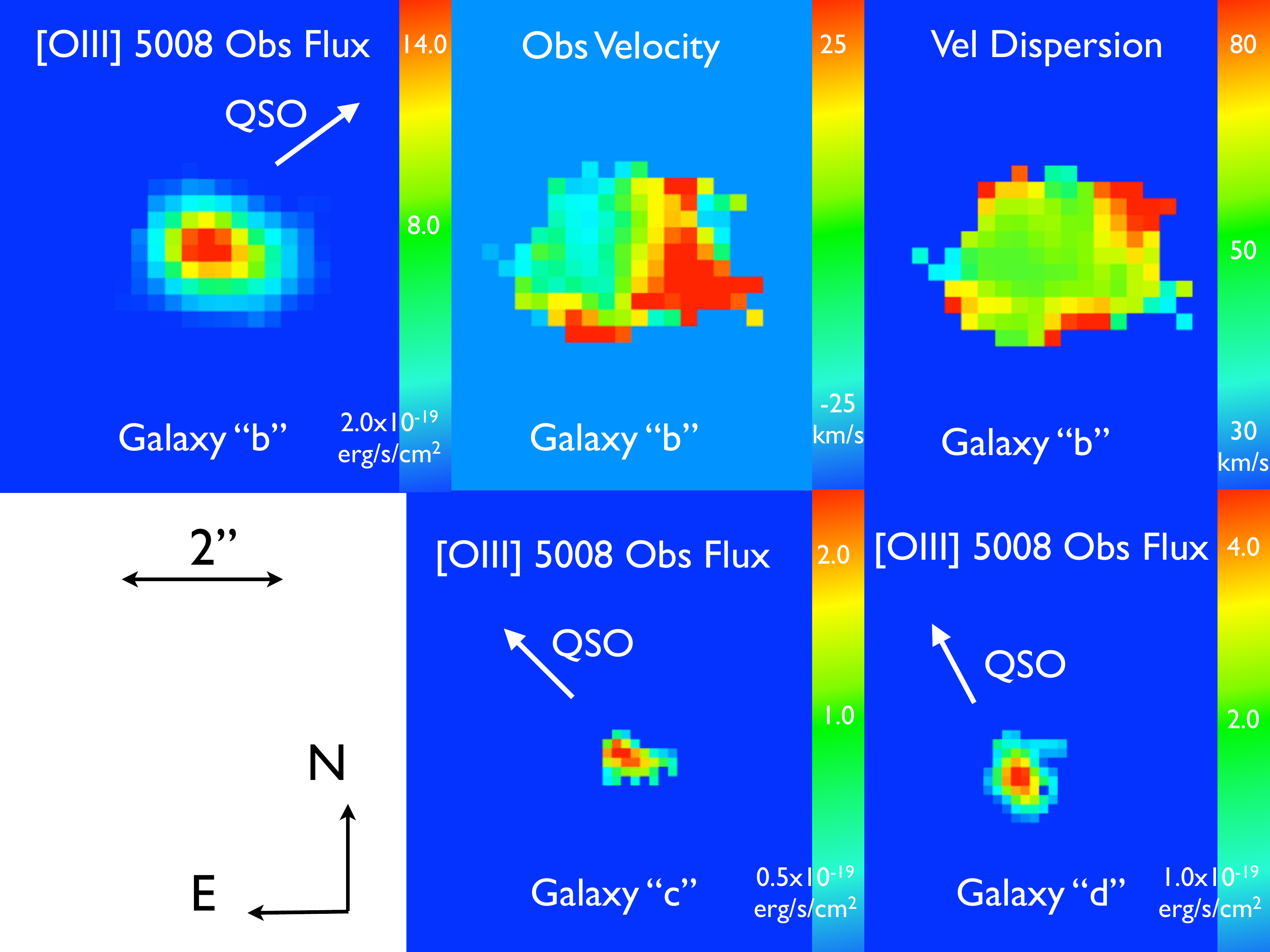}
\caption{{\bf Observed flux maps of galaxies "b", "c" and "d".} The top row shows the \oiii\ flux map, observed velocity map and dispersion map of galaxy "b". Although the object is compact, there are no indications of rotation from velocity and dispersion maps. Galaxies "c" and "d" \oiii\ flux maps are shown at the bottom. In all flux panels, the direction to the quasar is shown with a white arrow.}
\label{f:All_Gal_Kine}
\end{center}
\end{figure*}

\begin{table*}
\caption{{\bf Galaxy "a" morpho-kinematic properties.} These morphological and kinematic parameters are determined from our 3D fits using the GalPak$^{\rm 3D}$ algorithm. They are all corrected for inclination. PA stands for position angle. The azimuthal angle is the angle between the quasar line of sight and the projected major axis of the galaxy on the sky. The galaxy is inclined and presents all characteristics of a rotating disk. \label{t:Kine}}
\centering
\begin{tabular}{lccccccc}
\hline\hline
		&$r_{1/2}$ 	&sin $i$  		&P.A.  		&azimuthal angle& $V_{\rm max}$ &$M_{\rm dyn}$ &$M_{\rm halo}$ \\
		&[kpc]		& 			&[deg] 		&[deg]		   &[km~s$^{-1}$]  		   &[M$_{\odot}$] &[M$_{\odot}$] \\
\hline
Galaxy a   &7.9$\pm$0.1 &0.87$\pm$0.01 &65$\pm$1 	&12$\pm$1			   &200$\pm$3 	 	   &7.4$\pm$0.4$\times$10$^{10}$ &2.9$\pm$0.2$\times$10$^{12}$  \\
\hline\hline
\end{tabular}\par
\end{table*}

To estimate the kinematics of the galaxies, we use two different algorithms. First, we use the Camel code \citep{epinat09} which fits simultaneously several emission lines in the presence of a continuum. In this case, we fit \hb\ and doublets from \oii\ and \oiii\ simultaneously. In the case of galaxy "a", the continuum contribution comes from the galaxy continuum and the nearby quasar continuum. Second, we use the GalPak$^{\rm 3D}$ algorithm \citep{bouche15} which compares directly the data cube with a parametric model mapped in $x,y,\lambda$ coordinates. The algorithm uses a Markov Chain Monte Carlo (MCMC) approach with a non-traditional proposal distribution of the parameters in order to efficiently probe the parameter space. The code has the advantages that it fits the galaxy in 3-d space and provides a robust description of the morpho-kinematics of the data even in poor seeing conditions. We set the input parameters and model profiles and compare them with the actual data so as to obtain a convergence of the 10 free parameters (sky position, flux, half-light radius, inclination, PA, turnover radius, maximum velocity and velocity dispersion) over a 15,000 long MCMC in order to robustly sample the posterior probability distribution. Details of the method can be found in \citet{bouche15}.

The code provides stable results given certain conditions. First of all, the signal-to-noise (SNR) per spaxel of the brightest spaxel should be $>$3. For Galaxy "a", the SNR per spaxel of the brightest pixel in the \hb\ cube is 4.3. For Galaxy "b", it is only $\sim$2. Indeed, GalPak$^{\rm 3D}$ does not converge for galaxy "b" in part because the object is compact and also because the SNR per spaxel of the brightest spaxel is below the stability criterion for the algorithm convergence. Similarly, galaxies "c" and "d" are too faint (L/L$*$$\sim$0.01) and compact to perform a robust morpho-kinematics analysis. Second, \citet{bouche15} note that the algorithm performs well if the half-light radius is such that r$_{1/2}$/FWHM$>$0.75, which in our case (FWHM=0.72") translates into r$_{1/2}>$0.54" or r$_{1/2}>$2.7 pixel (0.2" pixel scale). We check that the r$_{1/2}$ value derived by the model satisfies this condition.
Third, we examined the covariance between the fitted parameters to check that potential degeneracies are broken (typically between V$_{\rm max}$ and inclination $i$ and V$_{\rm max}$ and turn-over radius r$_v$). To overcome the covariance between r$_v$ and V$_{\rm max}$, we set  the  turnover radius $r_v$ to satisfy the scaling relation between $r_v$ and the disk exponential $R_d$ found in local disk samples, which is approximately $r_{\rm t} \simeq R_d\times0.9$. In practice, we fixed r$_v$ to 1.5 kpc (i.e. $\sim$1.3 spaxel). Fourth, we aim at an acceptance rate (fraction of useful iterations count to total iterations count) of the MCMC between 30 and 50\%. In case of galaxy "a", we need to adjust the maximum number of accepted iterations ({\it random\_scale}) to reach such values of acceptance rate. Finally, we examined the residual maps to make sure that no significant features are present after the model is subtracted from the data. We use the default exponential flux profile but note that using a Gaussian flux profile does not affect the results. In all cases, we use an arctan velocity profile which best represent the data.

The [O\,\textsc{iii}]\,$\lambda 4960$ line was not used, but \oiii\ $\lambda5008$ \AA\ and \hb\ were fitted independently and the resulting fit parameters agree very well within the error estimates. We find that the maximum circular velocity is well constrained at around V$_{\rm max}$=200$\pm$3 km~s$^{-1}$. The half-light radius converges at about 7.9$\pm$0.1 kpc, and the position angle (PA) is found to be 65$\pm$1 deg. The inclination is sin $i$=0.87$\pm$0.01, i.e. slightly higher than the one reported by \citet{kacprzak11b} from a GIM2D modelling of HST observations (sin $i$=0.75$\pm$0.05). These results are collected in Table~\ref{t:Kine}.

The morphological analysis indicates the presence of spiral arms or possibly tidal tails in an interacting system as well as sub-structure in \oiii\ on scales of 1". The galaxy has a large inclination angle. Both the observed and model velocity maps indicate a gradient along the major axis of the galaxy. In addition, the dispersion peaks in the center of the object. All these characteristics point to a large rotating disk. We note that the kinematic model from \citet{muzahid16} would also be consistent with rotation had they used the redshift measured here z$_{\rm gal}$=0.43005. Indeed, our measure is probably more reliable as it is based on several high S/N emission lines while they have access to a rather noisy spectrum covering only \ha\ and \nii\ lines. Figure~\ref{f:Gal_a_Kine} shows the resulting \oiii\ flux and velocity maps, where the direction to the quasar is indicated with a white arrow. The maps in the bottom row of Figure~\ref{f:Gal_a_Kine} illustrate the results from our 3D modelling. The model is a good representation of the data, which can be appreciated from the residuals from the subtraction of the model from the observations shown in bottom-right panel. 

Given that galaxy "a" is rotating, we can use the enclosed mass to determine the dynamical mass within r$_{1/2}$ (Epinat et al. 2009):

\begin{equation}
M_{\rm dyn} = V_{\rm max}^2~r_{1/2}~/~G
\end{equation}

\noindent
where $V_{\rm max}$ and r$_{1/2}$ are measured from kinematic analysis above and are corrected for inclination. We therefore find $M_{\rm dyn}$=7.4$\pm$0.4$\times$10$^{10}$ M$_{\odot}$.

We are able to estimate the mass of the halo in which the systems reside, assuming a spherical virialised collapse model (Mo \& White 2002):

\begin{equation}
M_{\rm halo}= 0.1 H_0^{-1} G^{-1} \Omega_m^{-0.5} (1+z)^{-1.5} V_{\rm max}^3
\end{equation}

\noindent
using the V$_{max}$ value. We find  M$_{\rm halo}$=2.9$\pm$0.2$\times$10$^{12}$ M$_{\odot}$ for galaxy "a". 
These results are listed in Table~\ref{t:Kine}.

The top row of Figure~\ref{f:All_Gal_Kine} presents the \oiii\ flux map, observed velocity and velocity dispersion maps of galaxy "b". Although the object is compact, there are no indications of rotation from velocity and dispersion maps. Galaxies "c" and "d" \oiii\ flux maps are shown at the bottom of Figure~\ref{f:All_Gal_Kine}.

\section{Nature of the Gas}

In this study, we are able to combine information on the properties of neutral gas seen in absorption in front of the quasar Q2131$-$1207 with the wealth of data provided by MUSE on a group of galaxies at the same redshift as the absorber, including impact parameters, star formation rates, metallicities, morphological and kinematic properties. In the following, we review possible scenarios to explain the relation between the gas seen in absorption and the four galaxies we have identified. However, we consider galaxies "c" and "d" to be too far away from the quasar sightline (at $26.0^{\prime\prime}$ and $30.7^{\prime\prime}$ respectively, corresponding to 147\,kpc and 174\,kpc at the redshift of the sub-DLA) to be \textit{directly} associated with the gas observed in absorption (see e.g. \citet{peroux05} and \citet{rubin15} for an estimate of the size of quasar absorbers).

\subsection{Gas Rotating with the Disk}

Based on the observed velocity and dispersion maps (see Figure~\ref{f:All_Gal_Kine}), galaxy "b" does not show signs of rotation nor other significant velocity gradient. This could partly be an effect of beam smearing in natural seeing observations, limiting our ability to discern rotation in galaxy "b". In addition, it is slightly further away (61 kpc) than galaxy "a" (52 kpc) and with a larger velocity offset (z$_{\rm gal}$=0.43072) from the main component of the absorber, z$_{\rm abs}$=0.42980. For these reasons, we explore the possibility that the gas seen in absorption is related to the gas in rotation around galaxy "a", which lies at z$_{\rm gal}$=0.43005. In that object, we measure V$_{\rm max}$=200$\pm$3 km~s$^{-1}$ with the blueshifted side of the galaxy oriented towards the quasar line-of-sight (see Figure~\ref{f:Gal_a_Kine}). This is broadly consistent with what is seen in the absorption profile, with most of the absorption taking place at velocities between $v \simeq -100$ and $-50$\,km~s$^{-1}$ from the systemic redshift of galaxy "a" (see Figure~\ref{f:Gal_a_Kine}). In addition, the metallicity measured in absorption is comparable with the one expected from the measured flat internal metallicity gradient at impact parameter $\delta$=52 kpc.
However, we note that this scenario does not explain the weak components seen in absorption in the strong lines of \mgii\ and redwards of the systemic redshift of the galaxy at $v \simeq +50$\,km~s$^{-1}$. This further calls for additional physical processes to explain the gas seen in absorption.

\subsection{Gas Flows}

Simulations indicate that in galaxies driving large-scale outflows the outflowing gas preferentially leaves the
galaxy along its minor axis (path of least resistance), preventing infall of material from regions above and
below the disc plane (Brook et al. 2011). As a result, inflowing gas is almost co-planar with the galaxy disc
and is seen preferentially along the major axis (Stewart et al. 2011; Shen et al. 2012; Bordoloi et al. 2012).
Thus, information on orientation potentially allows one to observationally distinguish winds from accreting gas (e.g. Peroux et al. 2013; Schroetter et al. 2015). 

Because it is compact, we cannot constrain the inclination of galaxy "b", which prevents us from measuring its azimuthal angle. For galaxy "a", we measure of the azimuthal angle between the quasar line of sight and the projected 
major axis of the galaxy on the sky. We find the azimuthal angle to be 12$\pm$1 degrees, which would exclude the outflow scenario. This is further supported by the fact that the integrated metallicity of the galaxy, 12+log(O/H)=8.98$\pm$0.02, is higher than the metallicity of the absorbing gas, 12+log(O/H)=8.15$\pm$0.20. 

On the contrary, based on geometry and orientation arguments, it is possible that the gas seen in absorption is related to infalling cold gas onto galaxy "a". This is further supported by the metallicity difference between the galaxy and the sub-DLA absorber. The object lies in the infalling part of the galaxy to gas metallicity difference versus azimuthal angle plot proposed by \citet{peroux16}. However, with the present data, we cannot exclude that the gas is possibly (or also) infalling onto galaxy "b".

\subsection{Intra-group Gas}

Galaxies "c" and "d" are too far away from the quasar line-of-sight at 147 and 174 kpc (respectively) to be directly associated with the absorber and play a direct role in the possible scenarios above. However, their presence indicates that they are likely to trace a more general structure at that redshift. Given the separation between the galaxies which are the furthest apart in velocity space (galaxies "b" and "d"), we calculate that our observations probed a volume of 0.694 Mpc$^3$ at the redshift of the absorber. We use the luminosity function at this redshift reported by \citet{cowie10} to estimate that only 0.02 galaxies should be expected above our sensitivity limit (see also \citet{deharveng08} for the determination of the low-redshift luminosity function). Therefore, the four objects detected represent a significant overdensity. 

The two closest objects (galaxies "a" and "b") are $\delta$=11.3 arcsec (64 kpc) apart which means that they could be interacting, possibly tidally stripping. Given that the distance between the two objects with the largest separation on the plane of the sky (galaxies "a" and "d") is $\sim 200$\,kpc, the data at hand point to a group scale structure. However, the lack of information on larger scales prevent us from excluding a cluster. The velocity dispersion of the four objects is 81 km~s$^{-1}$ which is in the range of small galaxy groups. Based on such dispersion and assuming a virialised spherical system, we obtain a virial radius of 200 kpc. The virial mass for such system will be 3.3$\times$10$^{11}$M$_{\odot}$. This value is an order of magnitude lower than the halo mass for galaxy "a": M$_{\rm halo}$=2.9$\pm$0.2$\times$10$^{12}$M$_{\odot}$. However, the virial mass of the halo should be considered as a lower limit, since we may not have seen all group members, leading to underestimates of the dispersion and virial radius. In addition, we note the possible alignments of the objects in what could be a filament with a north-east to south-west orientation \citep{moller01, fumagalli16}.

\section{Discussion}

The large amount of data gathered in this work allows us to put constraints on possible scenarios explaining the origin of the gas seen in absorption. We find that part of the absorption profile can be explained by the rotating disk of galaxy "a". We exclude the possibility that the absorption is due to outflows but favour a scenario where some of the gas is infalling onto galaxy "a" and/or "b". Additionally, we argue that some gas may be related to the larger structure (group or filament) traced by the presence of four galaxies at the redshift of the absorber. We further note that several of these scenarios in combination might explain the final absorption profile. It is likely that neutral gas probed in absorption is related in part to gas bound to a large object in a group and/or filament and in part to intra-group gas. If that is the case, one can then question whether the physical properties of the main member of the group are a good representation of the galaxy properties seen in absorption, regardless of the exact identification of the absorbing gas. 

These results have several implications. First of all, this case and others \citep{bielby16, fumagalli16} challenge a possibly over-simplified picture where gas can be associated with a single isolated object and gas flow directions can be solely determined from geometry and orientation considerations. Indeed, this approach needs to be combined with information on the metallicity of the gas which is key to a complete understanding of gas flows. Only by selecting absorbers with measured $N$(H\,\textsc{i}) can one obtain a robust estimate of the absorbing gas metallicity as was the case here. While the current observations from both these indicators do not always paint a coherent picture \citep{peroux16}, it might also well be that only some of the components in the absorption profile can be explained by one or the other of the scenarios proposed above.

Second, it appears that studies at low-redshift provide a good strategy to reach a complete understanding of the relation of the absorbing gas to galaxies. Indeed, at redshift $z < 1$ current observations can access faint galaxies at the same redshift as the absorber down to L/L$*$$\sim$0.01, allowing a more comprehensive picture of the absorber environment to be assembled than is possible at $z > 1$. For the brightest member in the structure, detailed morpho-kinematic properties can be derived in a short observing time which would not be possible at higher redshifts. Finally, the MUSE optical coverage means that at z$<$0.8, the lines of \oii, \hb\ and \oiii\ are covered leading to robust redshift, SFR and metallicity estimates in emission.

\section{Conclusion}

The results presented in this work can be summarised as follows:

\begin{itemize}

\item We have obtained integral field spectroscopic observations with the VLT/MUSE instrument of a 
$\sim 1$\,arcmin field centred on the quasar Q2131$-$1207. The sightline to this quasar intersects a sub-DLA absorber at $z_{\rm abs} = 0.42980$, characterised by neutral hydrogen column density $\log [N$(H\,\textsc{i})]/cm$^{-2} = 19.50\pm0.15$ and abundance ${\rm [X/H]} = -0.54 \pm 0.18$. Approximately 0.1\% of the absorbing gas is in molecular form.

\item We identify four galaxies at redshifts consistent with that of the sub-DLA absorber, where only one (the brightest of the four) was known previously. The two galaxies closest to the quasar sight-line (52 and 61 kpc) exhibit signs of on-going star formation; while the other two, which are further away at projected distances of $> 140$\,kpc, are passive galaxies. We report the metallicities of the HII regions of the closest of these objects (12+log(O/H)=8.98$\pm$0.02 and 8.32$\pm$0.16) which are to be compared with the metallicity measured in absorption in the neutral phase of the gas (12+log(O/H)=8.15$\pm$0.20). For galaxy "a", we derive a flat metallicity gradient of +0.01$\pm$0.03 dex kpc$^{\rm -1}$.

\item For the brightest object, dubbed galaxy "a", a detailed morphological analysis indicates that the object is a extended galaxy showing indications of sub-structure on scales of 1". The kinematical modelling shows that it is an inclined (sin $i$=0.87$\pm$0.01) large rotating disk with V$_{\rm max}$=200$\pm$3 km~s$^{-1}$. We measure the dynamical mass of the object to be $M_{\rm dyn}$=7.4$\pm$0.4$\times$10$^{10}$ M$_{\odot}$ while the halo mass is M$_{\rm halo}$=2.9$\pm$0.2$\times$10$^{12}$ M$_{\odot}$.

\item Some of the gas seen in absorption is likely to be co-rotating with galaxy "a", possibly due to a warped disk. In addition, we measure an azimuthal angle of 12$\pm$1 degrees which may suggest that some fraction of the absorption may arise in gas being accreted \citep{peroux16}. This is further supported by the galaxy to gas metallicity difference, where the metallicity of the gas is found to be lower than the one from the galaxy.  We exclude outflows as a possibility to explain the gas in absorption but speculate that some of it may be related to intra-group gas or filament based on indications of a large structure (at least 200 kpc wide) formed by the four galaxies. 

\item Our new observations of the field indicate that the closest object to the quasar line-of-sight (a dwarf galaxy at $z_{\rm gal} = 0.74674$) is unrelated to the absorber. This chance alignment once more calls for caution when associating absorbers with galaxies without spectroscopic information.

\end{itemize}

Finally, it is clear from the work presented here that further MUSE observations of low-redshift (z$<$0.8) absorbers in quasar fields hold great potential for constraining the properties of the circumgalactic medium of galaxies.

\section*{Acknowledgements}
We would like to thank the referee for a constructive report. We are grateful to the Paranal and Garching staff at ESO for performing the observations in service mode and the instrument team for making a reliable instrument. We thank Michele Fumagalli for comments on an early version of the draft. We are grateful to Thomas Ott, Nicolas Bouch\'e and Jean-Charles Lambert for developing and distributing the QFitsView, GalPak$^{\rm 3D}$ and glnemo software, respectively. CP thanks the ESO science visitor programme for support and Wolfgang Kerzendorf for python tips. SM thanks CNRS and CNES (Centre National d'Etudes Spatiales) for support for his PhD. VPK acknowledges support from the U.S. National Science Foundation grant AST/1108830, and additional support from NASA grant NNX14AG74G and NASA Herschel Science Center grant
1427151. L.A.S. acknowledges support from ERC Grant agreement 278594-GasAroundGalaxies. This work has been carried out thanks to the support of the OCEVU Labex (ANR-11-LABX-0060) and the A*MIDEX project (ANR-11-IDEX-0001-02) funded by the "Investissements d'Avenir" French government program managed by the ANR.

\bibliographystyle{mn2e}
\bibliography{/Users/celine/Astro/Paper/MUSE_KnownPairsz0p4_Apr16/bibliography.bib}

\begin{appendix}

\section{MUSE Galaxy Spectra unrelated to the Quasar Absorber}

\begin{figure}
\includegraphics[angle=0,scale=0.43]{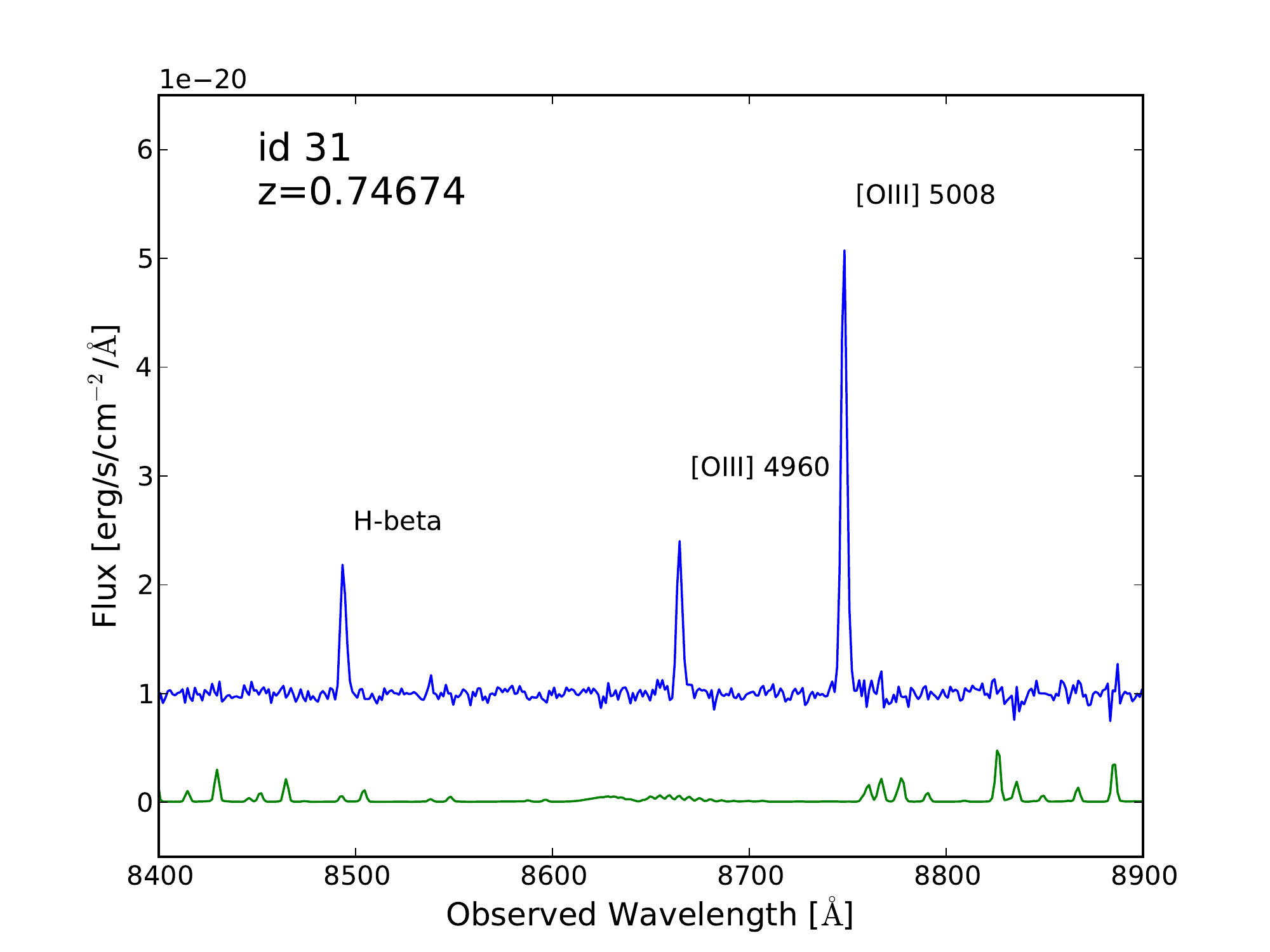}
\caption{{\bf MUSE spectrum of a dwarf galaxy}. The object lies at an impact parameter $\delta$=2.06" away from the quasar line-of-sight and is clearly detected in the HST/WFPC2 images (see Figure~\ref{f:hst}). The numerous emission lines are used to constrain the redshift to be z$_{\rm gal}$ =0.74674, i.e. higher than the redshift of the quasar.}
\label{f:NearObj}
\end{figure}

This appendix presents two MUSE spectra of objects unrelated to the quasar absorber at $z_{\rm abs} = 0.42980$. The first is a dwarf galaxy that is nearer to the quasar slight-line than any of the four galaxies discussed in main text (at impact parfameter
$\delta = 2.06^{\prime\prime}$). The object shows a number of strong emission lines including \oii, \neiii\ $\lambda$ 3871, H$\delta$, H$\beta$, \oiii\ at z$_{\rm gal}$=0.74674, i.e. higher than the redshift of the quasar. A portion of our the MUSE spectrum of that object is reproduced in Fig~\ref{f:NearObj}. 

Given the MUSE wavelength coverage, the most prominent absorption lines associated with the sub-DLA which are covered are the \caii\ H\&K lines. We recall that \citet{muzahid16} report log N(\caii)=11.98$\pm$0.10 based on the Keck/HIRES spectra. The doublet is also clearly visible in the MUSE spectrum of the background quasar. However, notwithstanding the high signal-to-noise of the nearby dwarf galaxy spectrum at these wavelengths, we do not detect the \caii\ doublet in absorption along this adjacent sightline, separated by 11.6\,kpc from the quasar sightline at the redshift of the sub-DLA. We calculate that given the quality of the spectrum we should only be able to measure N(\caii)$>$12.09 at 3$\sigma$, so that the non-detection is consistent with the higher resolution measurement in the quasar spectrum. We note that in the future such bright objects will be of use to probe the gas in absorption against multiple background sources.

\begin{figure}
\includegraphics[angle=0,scale=0.43]{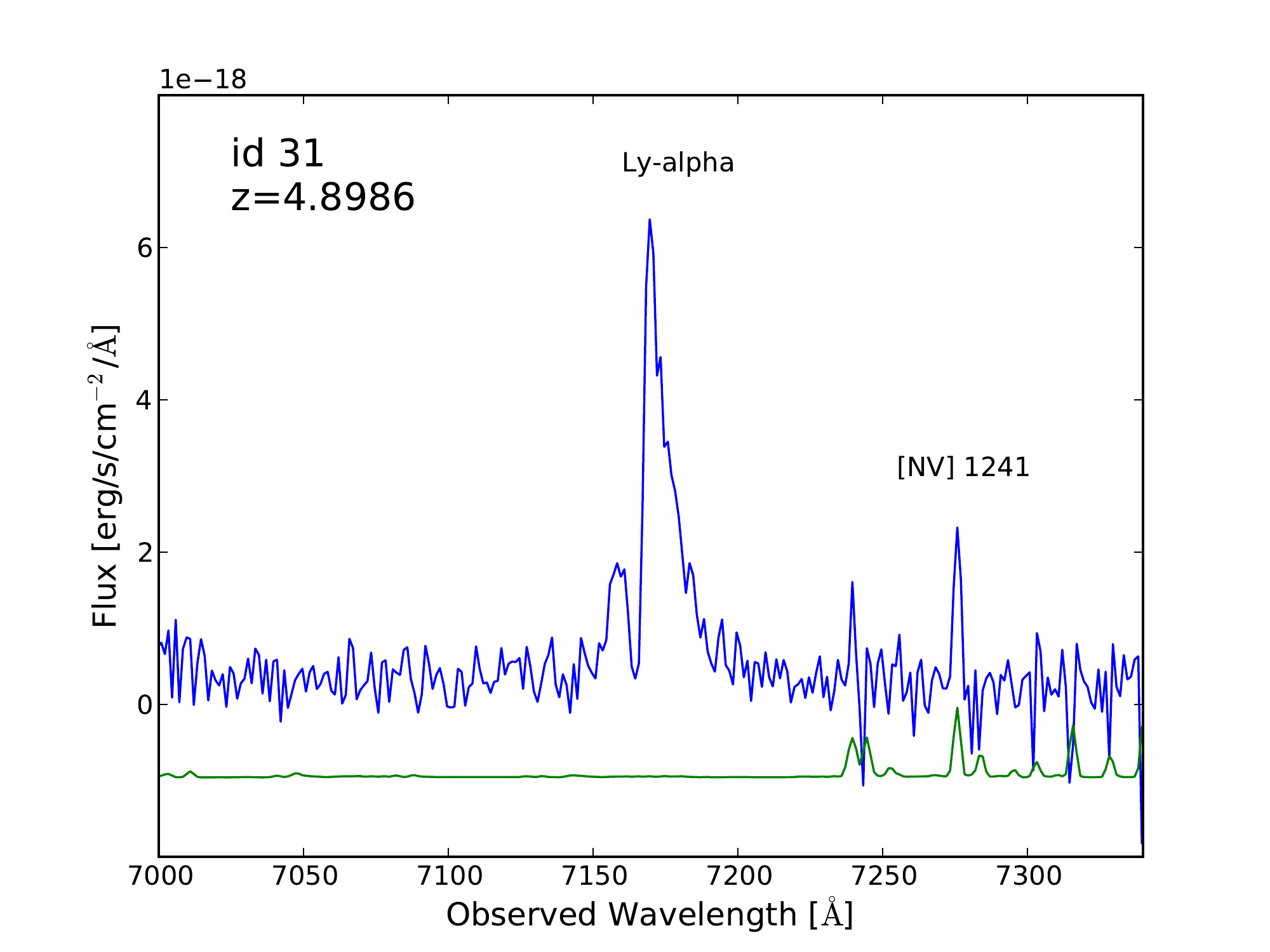}
\caption{{\bf MUSE spectrum of a strong emitter, coincident with the redshifted wavelength of \oiii\,$\lambda 5008$ at the redshift of the sub-DLA.}
The emission line shown is undoubtedly Ly$\alpha$ from a galaxy at $z_{\rm gal} = 4.8986$, as indicated by  the characteristic asymmetric emission + absorption line profile and the presence of narrow N\,\textsc{v}\,$\lambda 1240.81$ emission. This object is marked as "LAE" in Figure~\ref{f:NB} and shown with a red arrow in Figure~\ref{f:3Drendering}. }
\label{f:LyaObj}
\end{figure}

The second object appears to be coincident with the redshifted wavelength of \oiii\,$\lambda 5008$ at the redshift of the sub-DLA. However, the emission line is undoubtedly Ly$\alpha$ from a galaxy at $z_{\rm gal} = 4.8986$, as indicated by  the characteristic asymmetric emission + absorption line profile \citep{verhamme08} and the presence of narrow N\,\textsc{v}\,$\lambda 1240.81$ emission (Figure~\ref{f:LyaObj}).

\end{appendix}

\label{lastpage}
\end{document}